\documentclass[10pt,twocolumn,superscriptaddress]{revtex4-1}

\usepackage{amsmath, bm, amsfonts, amssymb}     
\usepackage{graphicx}                       
\usepackage{xfrac}                          
\usepackage{grffile}                        
\usepackage{color,colortbl}
\usepackage{xcolor}

\usepackage[colorlinks, citecolor=blue, urlcolor=red, linkcolor=blue]{hyperref}

\newcommand{\Bmax}{B_{\rm max}}
\newcommand{\Smax}{S_{\rm max}}

\newcommand{\gammabar}{\bar{\gamma}}
\newcommand{\gdot}[0] {\dot{\gamma}}
\newcommand{\gdotbar}[0] {\bar{\dot{\gamma}}}

\newcommand{\xhat}{\hat{\mathbf{x}}}
\newcommand{\yhat}{\hat{\mathbf{y}}}

\newcommand{\xg}{x_{\rm g}}
\newcommand{\tw}{t_{\rm w}}

\newcommand{\be}{\begin{equation}}
\newcommand{\ee}{\end{equation}}
\newcommand{\bea}{\begin{eqnarray}}
\newcommand{\eea}{\end{eqnarray}}

\begin{document}

\title{Ductile and brittle yielding in thermal and athermal amorphous materials}
\author{Hugh J. Barlow, James O. Cochran and Suzanne M. Fielding}
\affiliation{Department of Physics, Durham University, Science Laboratories,
  South Road, Durham DH1 3LE, UK}

\begin{abstract}

We study theoretically the yielding of sheared amorphous materials as
a function of increasing levels of initial sample annealing prior to
shear, in three widely used constitutive models and three widely
studied annealing protocols. In thermal systems we find a gradual
progression, with increasing annealing, from smoothly ``ductile''
yielding, in which the sample remains homogeneous, to abruptly
``brittle'' yielding, in which it becomes strongly shear banded. This
progression arises from an increase with annealing in the size of an
overshoot in the underlying stress-strain curve for homogeneous shear,
which causes a shear banding instability that becomes more severe with
increasing annealing.  ``Ductile'' and ``brittle'' yielding thereby
emerge as two limiting cases of a continuum of yielding transitions,
from gradual to catastrophic. In contrast, athermal systems with a
stress overshoot always show brittle yielding at low shear rates,
however small the overshoot.

\end{abstract}

\maketitle

Amorphous materials include soft glasses such as dense colloids,
emulsions, foams and
microgels~\cite{ISI:000407999000001,ISI:000419993800001,ISI:000266878700032,ISI:000246368400004,ISI:000293295000002},
as well as hard molecular and metallic
glasses~\cite{ISI:000374617600037,ISI:000320906700001}. Under low
loads or small deformations, such materials show solid-like
behaviour. Under higher loads or larger deformations, they yield
plastically. For some systems, the dynamical process whereby an
initially solid-like sample yields to give a finally fluidised flow is
rather smooth and
gradual~\cite{ISI:000295085700080,ISI:000277945900061,ISI:000301801100015,ISI:000332461800012,ISI:000268689400009,ISI:000280140800011,ISI:000261891200077,ISI:000365222200015,ISI:000357577400001}.
Others instead yield abruptly, with catastrophic sample
failure~\cite{schuh2007mechanical}.  For both ``ductile'' materials,
which yield smoothly and gradually, and ``brittle'' materials, which
yield abruptly and catastrophically, understanding the statistical
physics of yielding is the focus of intense current interest.  Theories have
been put forward based on a first order transition in a replica
theory~\cite{ISI:000370815100008,ISI:000410885300004}; a critical
point in an elastoplastic
model~\cite{ISI:000362909100023,liu2018creep}; a directed percolation
transition~\cite{ISI:000386386400004,ISI:000384392300003}; and a
spinodal~\cite{ISI:000309611400031,ISI:000402296700034,ISI:000346387700013,rainone2015following,urbani2017shear}. Microscopic
precursors to yielding have recently been observed in soft
materials~\cite{ISI:000429012500051,ISI:000392096800036,ISI:000341025700007}.

\begin{figure}[!b]
\vspace{-0.35cm}
\includegraphics[width=9.0cm]{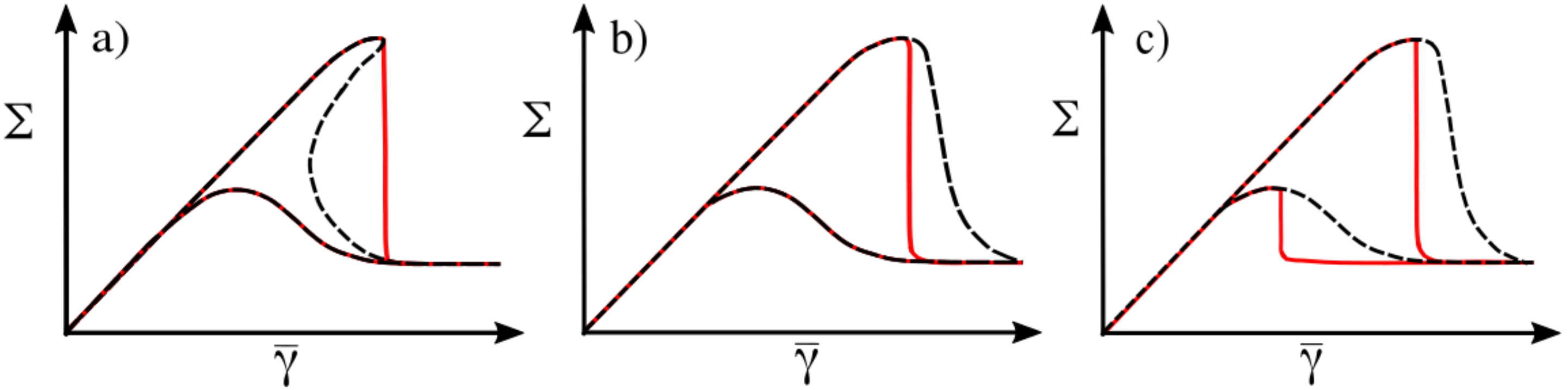}
\caption{Schematic  shear stress {\it vs.} strain  in different possible scenarios for  ductile and brittle yielding:  as suggested {\bf (a)} in Ref.~\cite{ISI:000436245000061,popovic2018elastoplastic}, and here for {\bf (b)} thermal  and  {\bf (c)} athermal systems. In each case the upper(lower) curve is for a  better(less well) annealed sample. Dashed lines: theoretical curve with homogeneous shear artificially enforced. Solid lines: precipitous drop from (a) stress overhang, or (b,c) from homogenous curve due to shear banding, giving ``brittle'' yielding.}
\label{fig:sketch}
\end{figure}

Recent mean field
calculations~\cite{ISI:000436245000061,popovic2018elastoplastic} have
suggested that the underlying stress-strain relation,
$\Sigma(\gamma)$, for an athermal amorphous material undergoing
quasistatic shear displays a qualitative change in form from the
lower to the upper curve in Fig.~\ref{fig:sketch}a) as the degree to
which a sample is annealed prior to shear increases.  Poorly annealed
samples (lower curve) then yield in a smoothly ``ductile'' way. Well
annealed samples (upper curve) instead show catastrophic ``brittle''
yielding, as the stress drops precipitously once the overhang is
reached.  In this scenario, ``ductile'' and ``brittle'' yielding are
separated by a random critical point, at which the stress-strain curve
switches between two qualitatively different shapes with increasing
annealing. Particle simulations were argued to agree with this
scenario~\cite{ISI:000436245000061}.

Here we propose an alternative scenario, in which the underlying
stress-strain curve for homogeneous shear has an overshoot, rather
than an overhang, followed by a regime of negative slope,
$\partial_\gamma\Sigma <0$. In thermal systems, where $k_{\rm B}T$ is
important ({\it e.g.} compared to local energy barriers for particle
rearrangements), a state of initially homogeneous shear becomes
linearly unstable to the formation of shear bands in this negatively
sloping
regime~\cite{ISI:000315141600016,ISI:000251326200013,ISI:000262976900016},
with the severity of banding (defined below) after an
(adimensionalised) stress drop of magnitude $\Delta\Sigma$ from the
stress maximum scaling as $\exp(\Delta\Sigma)$ in the limit of slow
shear. A large stress drop (strong annealing) thus causes severe
banding and correspondingly brittle failure, whereas a small stress
drop (weak annealing) causes only weak banding, and failure remains
ductile: Fig.~\ref{fig:sketch}b). In slowly sheared athermal systems,
in contrast, the severity of banding diverges (at the level of a
linear instability calculation) as the overshoot is approached as
$1/\partial_\gamma\Sigma$. Slowly sheared athermal systems accordingly
show brittle failure for any size of stress drop, however small:
Fig.~\ref{fig:sketch}c). We substantiate these scenarios both
analytically and numerically in a thermal fluidity
model~\cite{ISI:000286879900011} and an athermal elastoplastic
model~\cite{nicolas2018deformation}; and numerically~\cite{PRL_SI} in
the soft glassy rheology model~\cite{ISI:A1997WM06400048} in both
thermal and athermal regimes. Crucially, in neither thermal nor
athermal scenario does a critical point of the kind in
Fig.~\ref{fig:sketch}a) separate ductile and brittle yielding.

For consistency with the vocabulary adopted in
Ref.~\cite{ISI:000436245000061}, we use the term ``brittle'' to
characterise abrupt yielding in which the rate of failure is much
larger than the rate of the imposed deformation, and in which the
strain becomes strongly localised within the sample, but with a
caveat that we quote from~\cite{ISI:000436245000061}: ``Although this
phenomenon is not accompanied by the formation of regions of vacuum,
as it happens in the fracture of brittle materials, the macroscopic
avalanche taking place at the discontinuous yielding transition does
resemble a crack induced by a brittle fracture''.  Indeed, our
calculations and those of Ref.~\cite{ISI:000436245000061} are
performed at fixed volume, disallowing the opening of an air gap. We
suggest, however, that the formation of a severe high shear band in
abrupt yielding will, in studies at fixed pressure, indeed lead to the
rapid opening of an air gap.

We consider a sample prepared by some time $t=0$ with some level of
annealing (defined below) then sheared for all $t>0$ between infinite
flat parallel plates at $y=0, L_y$ by moving the top plate at speed
$\gdotbar L_y$ in the positive $\xhat$ direction. The shear is assumed
incompressible and inertialess.  As is standard practice, we restrict
all velocities $v(y,t)$ to the main flow direction $\xhat$, and
gradients to $\yhat$. The local shear rate $\gdot(y,t)=\partial_y
v(y,t)$, may vary across $y$ due to shear banding.  The spatially
average imposed shear rate $\gdotbar=\int_0^{L_y}dy\gdot(y,t)/L_y$. We
track only the shear component of the stress,
$\Sigma_{xy}(t)=\sigma_{xy}(y,t)+\eta\gdot(y,t)$, with an
elastoplastic contribution $\sigma_{xy}(y,t)$ and a Newtonian solvent
contribution of viscosity $\eta$. Force balance requires
$\partial_y\Sigma_{xy}=0$. Hereafter we drop the $xy$ subscript for
clarity. For the dynamics of the elastoplastic stress $\sigma$, we
consider three different constitutive models: a thermal continuum
fluidity model~\cite{ISI:000286879900011}, an athermal elastoplastic
model~\cite{nicolas2018deformation} and the soft glassy rheology
model~\cite{ISI:A1997WM06400048}, separately in thermal and athermal
regimes ~\cite{PRL_SI}.

{\it Thermal systems --} As a simple model of thermal systems we
consider a continuum fluidity model~\cite{ISI:000286879900011}, which
supposes a Maxwell-type constitutive equation:
\be
\partial_t\sigma(y,t)=G\gdot-\sigma/\tau.
\label{eqn:sigma}
\ee
Here $G$ is a constant modulus and $\tau$ is a stress relaxation time, which has its own dynamics:
\be
\partial_t\tau(y,t)=1-\frac{|\gdot|\tau}{1+|\gdot|\tau_0}+\frac{l_o^2}{\tau_0}\partial^2_y \tau.
\label{eqn:tau}
\ee
The first term on the RHS captures ageing, in which the timescale for
stress relaxation increases linearly with the time $\tw$ for which a
sample is aged before any deformation commences. The second term
captures rejuvenation by deformation, with $\tau_0$ a microscopic
timescale that sets the limiting value for $\tau$ as $\gdot\to\infty$.
The mesoscopic length $l_o$ describes the tendency for the relaxation
time of a mesoscopic region to equalise with its neighbours.

We consider a sample aged (annealed) for a time $\tw$ before shear
commences at time $t=0$, such that $\tau(y,t=0)=\tw$.  To seed
heterogeneity, we add noise in each numerical timestep $Dt$ as
$\sigma(y,t+Dt)=\sigma(y,t)+r\delta\sqrt{Dt}\cos(\pi y/L_y)$, with $r$
chosen from a top hat distribution between $-0.5$ and $+0.5$, and
$\delta$ small.  In both this model and the athermal one below, we
rescale stress, time and length so that $G=\tau_0=L_y=1$.  The solvent
viscosity $\eta\ll G\tau_0=1$ is unimportant to the physics we
describe. We use typical values $\eta=0.01-0.05$, but find no changes
to our results on reducing $\eta$ further.

\begin{figure}[!t]
\includegraphics[width=9.0cm]{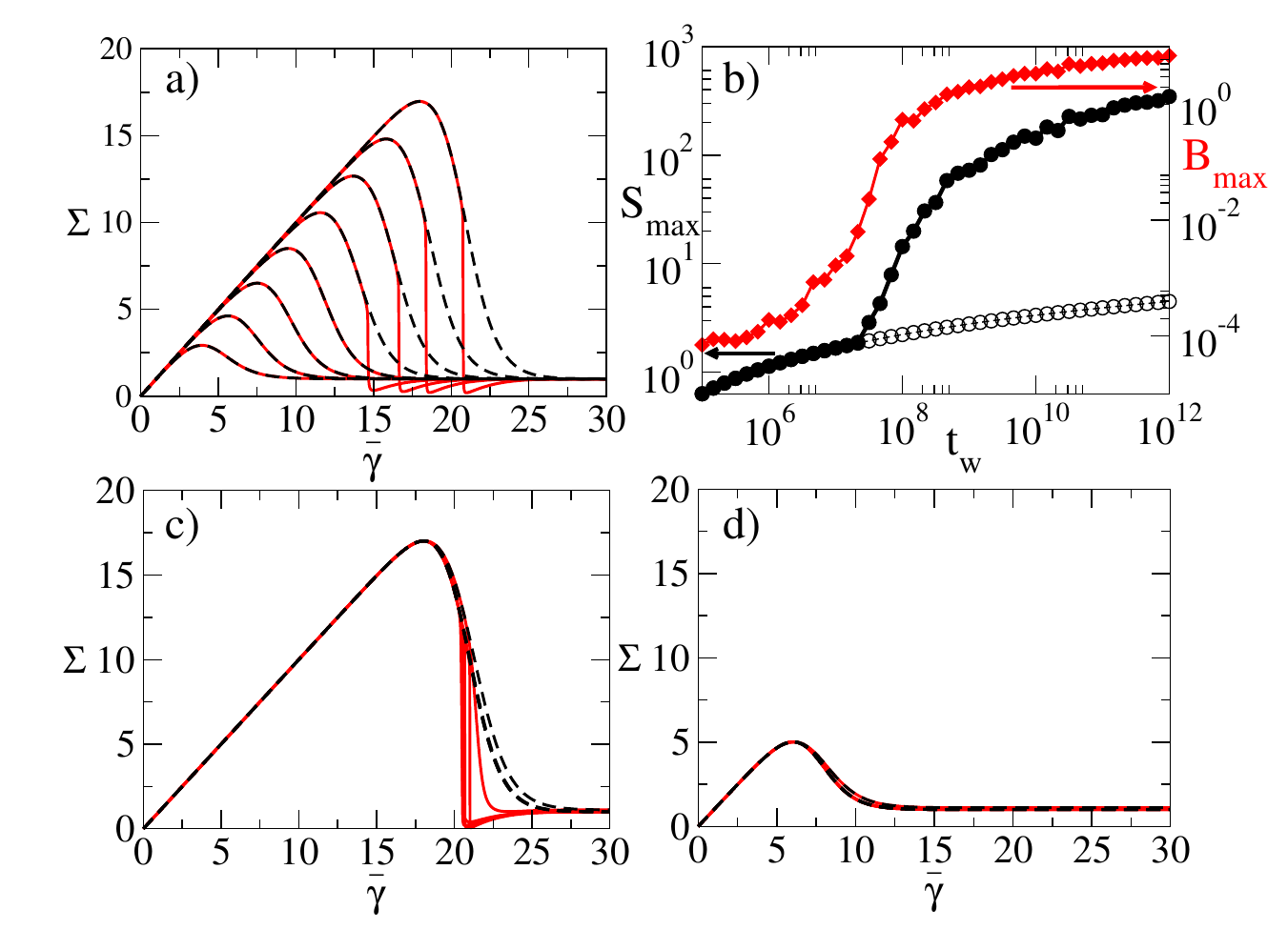}
\caption{{\bf a)} Stress {\it vs.} strain in thermal fluidity model
with homogeneous flow enforced (dashed lines) and shear banding
allowed (solid  lines). Imposed shear rate $\gdotbar=10^{-3}$, waiting
times $\tw=10^5,10^6\cdots10^{12}$  in curves left to right. {\bf b)}
Left vertical axis: steepest negative slope in stress-strain curve
with homogeneous flow enforced (open circles) and allowing banding
(closed black circles) {\it vs.} sample age $\tw$, for fixed $\gdotbar=10^{-3}$.
Right vertical axis: corresponding maximum degree of shear
banding. ($\Smax$ and $\Bmax$ each averaged over $10 - 60$ runs at
each $\tw$.)  {\bf c)} and {\bf d)} show curves in the same format as
{\bf a)}, but now for a fixed large stress peak $\Sigma_{\rm
max}=17.0$ {\bf (c)} or small stress peak $\Sigma_{\rm max}=5.0$ {\bf
(d)}, for imposed $\gdotbar=10^{-5},10^{-4}\cdots 10^{-1}$. Steeper
stress drop for smaller $\gdotbar$ in c). $\eta=0.05,
\delta=0.01\gdotbar, l_0=10^{-3},Dt=0.01,Dy=1/3000$.}
\label{fig:stress}
\end{figure}

Fig.~\ref{fig:stress}a) shows as dashed lines the stress
$\Sigma(\gammabar)$ as a function of the accumulating strain
$\gammabar=\gdotbar t$, calculated by imposing that the shear must
remain homogeneous across the sample, for several values of the sample
age $\tw$ at a single value of the imposed shear rate $\gdotbar$. Each
curve shows an initially solid-like elastic regime in which the stress
increases linearly with strain. At late times (large strains), the
sample flows plastically with a constant value of the shear
stress. For intermediate times (and strains), the stress shows an
overshoot that increases in amplitude with increasing degree of sample
annealing prior to shear (increasing $\tw$): an older sample shows a
larger initial regime of elastic response before yielding into a
finally flowing state.

We then performed separate calculations in which shear bands are
allowed to form. The resulting stress-strain curves are shown by solid
lines in Fig.~\ref{fig:stress}a). For poorly annealed samples, each
curve still follows that of the corresponding homogeneous calculation,
to good approximation, indicating that the shear field remains
homogeneous (or nearly so), with yielding occurring in a smoothly
gradual (``ductile'') way.  For well annealed samples, in contrast,
the stress-strain curve of the heterogeneous calculation only follows
that of the heterogeneous one until just after the stress
overshoot. It then drops precipitously as the sample becomes strongly
shear banded, causing abrupt (``brittle'') yielding.

\begin{figure}[!t]
\includegraphics[width=9.0cm]{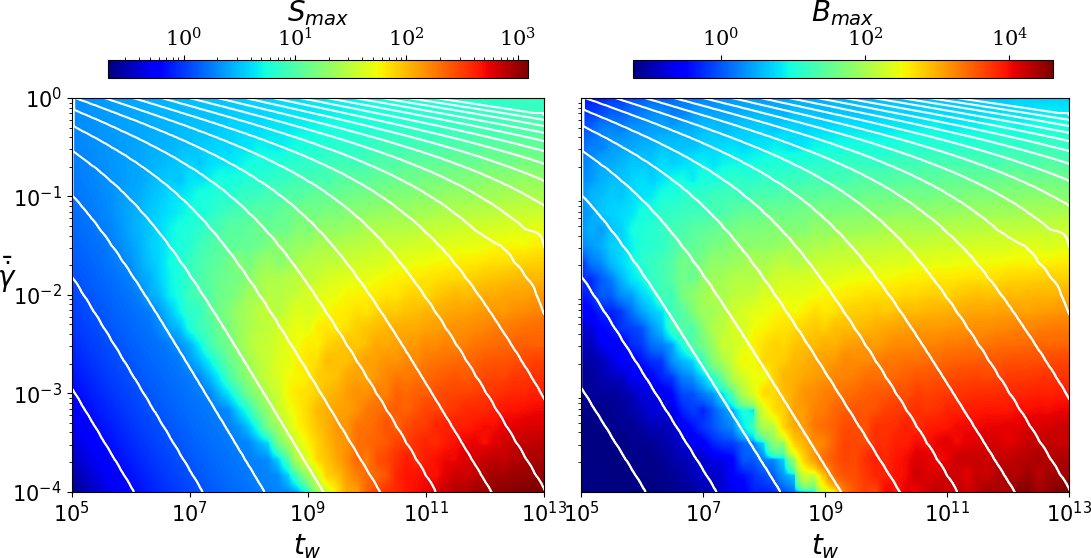}
\caption{Colourscale showing in thermal fluidity model {\bf left)}  steepest negative slope of stress versus strain, $\Smax$, and {\bf right)}  maximum degree of shear banding, $\Bmax$, during deformation. Each coordinate pair in the plane represents an average over $10-60$ deformation  simulations, each with an initial sample age $\tw$ and imposed shear rate $\gdotbar$. Contour lines show size of stress overshoot (peak minus steady state stress) $2, 4, 6,\cdots 40$ (bottom  left to top right). $\eta=0.05, \delta=0.01\gdotbar, l_0=10^{-3}, Dt=0.01, Dy=1/3000$.}
\label{fig:phaseDiagrams}
\end{figure}

For each individual deformation experiment, defined by the values of
$(\gdotbar,\tw)$, we denote by $\Smax$ the absolute value of the
maximally negative value of $\partial_\gamma\Sigma$ (with the
maximisation performed over all times during the deformation). This
quantifies the abruptness of the stress drop during yielding. We
further denote by $\Bmax$ the degree of shear banding, again maximised
over all times during the deformation.  (At any time $t$, or strain
$\gammabar(t)$, we define the degree of shear banding $B(\gammabar)$
as the maximum minus minimum local strain rate
across the sample, normalised by $\gdotbar$.)  This quantifies the severity of shear banding
during yielding. These quantities are plotted as a function of sample
age in Fig.~\ref{fig:stress}b), again for a fixed $\gdotbar$. A regime
of gradual yielding (low $\Smax$) and near homogeneous deformation
(low $\Bmax$) for poorly annealed samples (low $\tw$) crosses over
into a regime of precipitous yielding (high $\Smax$) and strong
banding (high $\Bmax$) at high $\tw$. This crossover is explored in
the full plane of $\gdotbar,\tw$ in Fig.~\ref{fig:phaseDiagrams}. For
$\gdotbar
\ll 1$ it occurs at $\gdotbar\approx 10^{m}/\tw$, with $m\approx 4.5$. Deviations
from this scaling at higher $\gdotbar$ should be disregarded, because
the model is itself only valid for $\gdotbar\ll 1$. Increasing strain
localisation~\cite{ISI:000262976900016} and decreasing notch fracture
toughness~\cite{vasoya2016notch,ketkaew2018mechanical} with decreasing
initial `effective temperature' have been seen in the STZ model.

To understand these results, we perform a linear stability analysis
for how strongly shear bands will form during any deformation
experiment, by writing the system's state as the sum of a
time-dependent homogeneous base state (as would pertain in a
theoretically idealised deformation in which shear banding is
prohibited), plus an initially small heterogeneous precursor to any
shear bands: $\gdot(y,t)=\gdotbar+\delta\gdot(t)\exp(iky)$,
$\sigma(y,t)=\bar{\sigma}(t)+\delta\sigma(t)\exp(iky)$,
$\tau(y,t)=\bar{\tau}(t)+\delta\tau(t)\exp(iky)$. Substituting these
into the model equations and expanding to first order in the amplitude
of the heterogeneity, we find~\cite{PRL_SI} the degree of shear
banding $\delta\gdot(t)/\gdotbar$ after a stress drop of magnitude
$\Delta\Sigma$ from the stress overshoot to scale as
$(\delta\gdot_0/\gdotbar)\exp(\Delta\Sigma)$ in the limit $\gdotbar\to
0$, with $\delta\gdot_0/\gdotbar$ the small initial heterogeneity due
to noise. Systems with a small stress overshoot therefore remain
almost homogeneous and show ``ductile'' yielding, whereas those with a
large stress overshoot will show strong shear banding and ``brittle''
failure. ``Ductile'' and ``brittle'' yielding thus emerge as limiting
cases, arising from a smooth variation in the height of the stress
overshoot, which scales as a function of $\gdotbar\tw$ for low
$\gdotbar$ (contour lines in
Fig.~\ref{fig:phaseDiagrams}). Deformation experiments with a high
$\gdotbar\tw$ thus show brittle failure at low $\gdotbar$ (red region
in Fig.~\ref{fig:phaseDiagrams} and low $\gdotbar$ curves in
Fig.~\ref{fig:stress}c); those with low $\gdotbar\tw$ show ductile
yielding for all $\gdotbar$ (dark blue region in
Fig.~\ref{fig:phaseDiagrams}; and Fig.~\ref{fig:stress}d).

In Fig. 1 of Ref.~\cite{PRL_SI}, we explore ``ductile'' and
``brittle'' yielding in the soft glassy rheology model in its thermal
regime. Each subpanel a)-d) of that figure is strikingly analogous to
its counterpart in Fig.~1, demonstrating the same scenario as in this
thermal fluidity model.

{\it Athermal systems --} As a simple model of an athermal amorphous
material we consider an ensemble of elastoplastic elements, each
corresponding to a local mesoscopic region of
material\cite{nicolas2018deformation}.  Given a shear rate $\gdot$,
each element builds up a local elastic shear strain $l$ according to
$\dot{l}=\gdot$, giving a local shear stress $Gl$ and energy
$\tfrac{1}{2}Gl^2$, where $G$ is a constant modulus.  This stress is
intermittently released by local plastic yielding events, which occur
stochastically with rate $r(l)=\tau_0^{-1}$ when a local energy
barrier $E$ is exceeded, $\tfrac{1}{2}Gl^2>E$, and $r(l)=0$ otherwise,
with $E=1$ (in our units) for all elements. Here $\tau_0$ is a
microscopic attempt time. Upon yielding, any element resets its local
stress to zero. The probability distribution $P(l,t)$ of local strains
obeys:
\be
\dot{P}(l,t)+\gdot\, \partial_l P=-r(l)P+Y(t)\delta(l).
\ee
Here $Y(t)=\int dl\, r(l)\,P(l,t)$ is the ensemble average local
yielding rate and  $\delta(l)$ is the Dirac delta function.  The total
elastoplastic stress $\sigma(t)=G \int dl\, l\, P(l,t)$.

So far, we have assumed homogeneous flow in this elastoplastic model,
without accounting for spatial stress propagation following any local
yielding event. To account for non-uniform shear deformations, we now
take $n=1...N$ streamlines at discretised flow-gradient positions
$y=0...L_y$, with periodic boundary conditions. The distribution
$P(l,y,t)$ on any streamline then obeys:
\be
\dot{P}(l,y,t)+\gdot(y,t)\, \partial_l P=-r(l)P+Y(y,t)\delta(l),
\ee
with streamline yielding rate $Y(y,t)=\int dl\, r(l)\,P(l,y,t)$ and
elastoplasic stress $\sigma(y,t)=G \int dl\, l\, P(l,y, t)$.  Given an
imposed average shear rate $\gdotbar$ across the sample as a whole,
the shear rate on each streamline is calculated by enforcing force
balance: $\sigma(y,t)+\eta\gdot(y,t)=\bar\sigma(t)+\eta\gdotbar$, with
$\bar\sigma(t)=\tfrac{1}{L_y}\int dy \sigma(y,t)$.  This ensures 1D
stress propagation on a timescale $\eta/G$ following any local
yielding event, recovering the 1D projection of the Eshelby propagator
of 2D lattice elastoplastic models~\cite{nicolas2018deformation}.

\begin{figure}[!t]
\includegraphics[width=9.0cm]{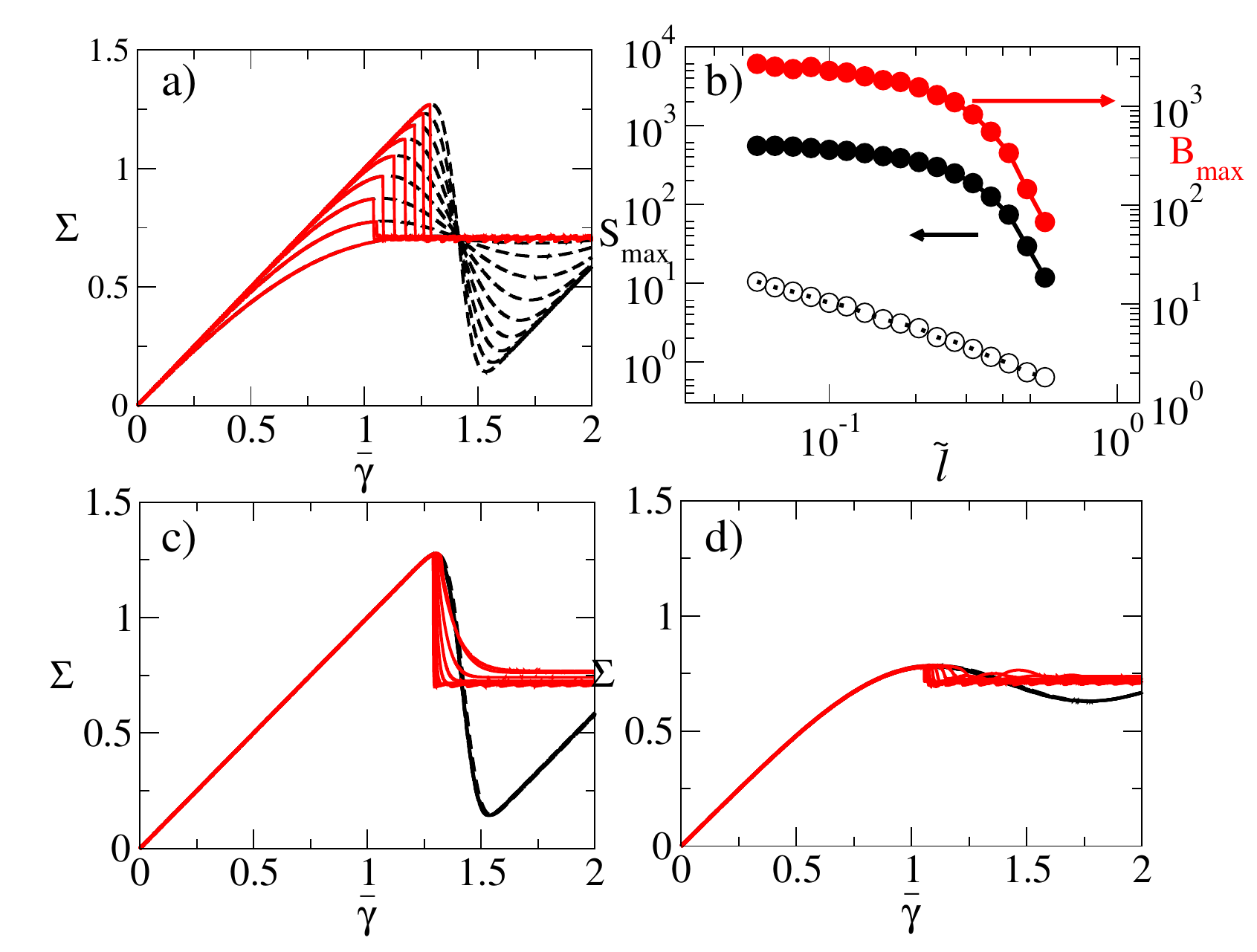}
\caption{{\bf a)} Stress {\it vs.} strain in athermal elastoplastic model
with homogeneous flow enforced (dashed lines) and shear banding
allowed (solid  lines). Shear rate $\gdotbar=10^{-4}$.
Annealing increases with decreasing $\tilde{l}=0.562,0.421,0.316,0.237,0.177,0.133,0.1,0.0749,0.0562$  in curves left
to right. {\bf b)} Left vertical axis: steepest negative slope in
stress-strain curve with homogeneous flow enforced (open symbols) and
allowing banding (closed black symbols) {\it vs.}  $\tilde{l}$, for fixed
$\gdotbar=10^{-4}$.  Right vertical axis: corresponding maximum degree
of shear banding.  {\bf c)} and {\bf d)}: curves in same
format as {\bf a)}, but now for fixed strong annealing $\tilde{l}=0.0562$ {\bf (c)} or  weak annealing $\tilde{l}=0.421$ {\bf
(d)}, for imposed $\gdotbar=10^{-n}$ with $n=2.00,~2.25\cdots 4.00$; steeper
stress drop for smaller $\gdotbar$. $\eta=0.05, w=0.05, N=20 ,M=80000, Dt=0.05$.}
\label{fig:EPMstress}
\end{figure}

We simulate this model by evolving $M=80000$ elastoplastic elements on
each of $N=20$ streamlines, with force balance across streamlines as
described.  Adjacent streamlines are further weakly coupled by
adjusting the stress of three randomly chosen elements on each
adjacent streamline an amount $wl(-1,+2,-1)$ following any yielding
event of size $l$, with $w$ a small parameter, mimicking the stress
diffusion term in Eqn.~\ref{eqn:tau} above.

Before shear commences at time $t=0$ we assign each element an initial
local strain from a Gaussian,
$P_0(l)=\exp(-l^2/2\tilde{l}^2)/\sqrt{2\pi}\tilde{l}$. Smaller values
of $\tilde{l}$ correspond to better annealed samples. Accordingly, we
characterise the degree of annealing by $1/\tilde{l}$, which forms the
counterpart to $\tw$ for thermal systems above.

Fig.~\ref{fig:EPMstress}a) shows as dashed lines the stress
$\Sigma(\gammabar)$ as a function of accumulating strain
$\gammabar=\gdotbar t$, calculated by imposing that the shear must
remain homogeneous, for several levels of annealing prior to shear,
$1/\tilde{l}$, at a single imposed $\gdotbar$. Each curve shows an
initially solid-like elastic regime in which the stress increases
linearly with strain, before the stress declines as plastic yielding
sets in. (The stress later shows a persistent oscillation known to
arise in homogeneous deformation of simplified elastoplastic
models~\cite{nicolas2018deformation}.) As in thermal systems, more
strongly annealed samples show a larger initial elastic regime.

We then performed separate calculations in which shear bands can
form. The resulting stress-strain curves are shown by solid lines in
Fig.~\ref{fig:EPMstress}a). As the stress overshoot is reached, the
stress in these heterogeneous calculations falls precipitously below
that of the homogeneous calculations as shear bands form, leading to
``brittle'' failure. By comparing each of
Figs.~\ref{fig:EPMstress}a,c,d) with its counterpart subpanel in
Fig.~\ref{fig:stress}, we see an important difference between thermal
and athermal systems. For thermal systems, a large enough stress
overshoot is required to see ``brittle'' failure. In contrast,
athermal systems show ``brittle'' failure however small the overshoot.
The trend to increasingly sharp failure with decreasing shear rate has
been demonstrated in particle simulations of a stable glass with a
large stress overshoot~\cite{singh2020brittle}.

\begin{figure}[!t]
\includegraphics[width=8.5cm]{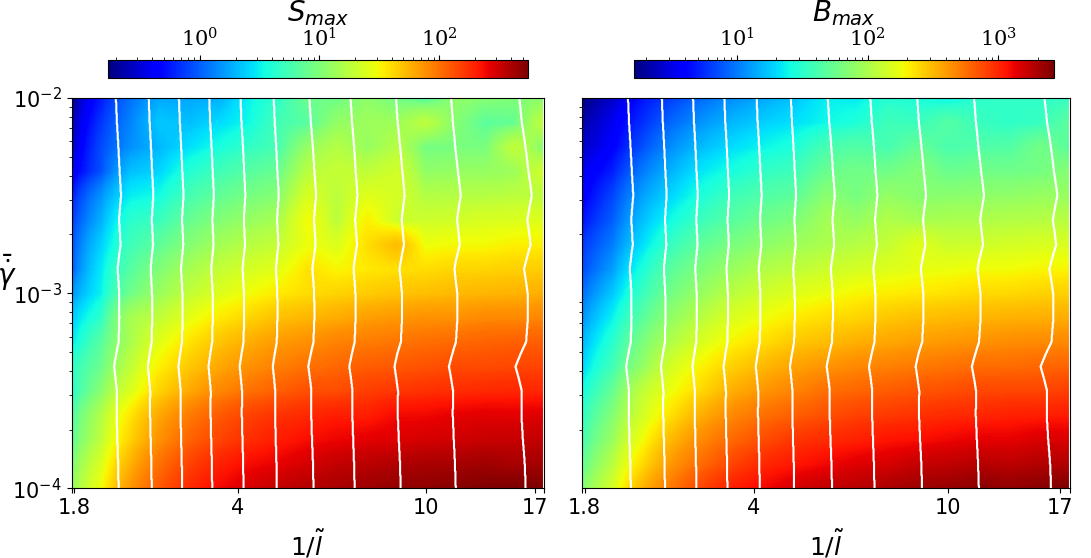}
\caption{Colourscale showing in athermal elastoplastic model {\bf left)}  steepest negative slope of stress versus strain, $\Smax$, and {\bf right)}  maximum degree of shear banding, $\Bmax$, during deformation. Each coordinate pair represents a deformation simulation with annealing parameter $1/\tilde{l}$ and imposed shear rate $\gdotbar$. Contour lines show size of stress overshoot (peak minus steady state stress) $0.05,0.10\cdots 0.55$ from left to right. $\eta=0.05, w=0.05, N=20 ,M=80000, Dt=0.05$.}
\label{fig:EPMphaseDiagrams}
\end{figure}

To explore this further, we show in Fig.~\ref{fig:EPMphaseDiagrams}
colourmaps of the maximum steepness of stress drop, $\Smax$, and
severity of shear banding, $\Bmax$, as a function of the degree of
sample annealing $1/\tilde{l}$ and imposed shear rate $\gdotbar$. (We
define the degree of banding at any time as the variance in shear rate
across the sample, then maximise this quantity over the deformation
simulation.) Increasing overshoot heights are shown by contour lines
left to right. Even for the smallest accessible overshoot height,
$\Bmax$ and $\Smax$ increase without bound as $\gdotbar\to 0$ in this
athermal system,  leading to ``brittle'' yielding.

To understand these results for athermal systems, we perform in
Ref.~\cite{PRL_SI} a linear stability analysis for how strongly shear
bands will form during any deformation experiment. In slow shear,
$\gdotbar\to 0$, we find the degree of shear strain banding
$\delta\gamma(t)$ to diverge as $\delta\gamma(0)/\partial_\gamma\Sigma$
on approach to the stress overshoot,
$\partial_\gamma\Sigma=0$. Athermal systems thus show ``brittle''
failure however small the stress overshoot, as indeed seen numerically
in Figs.~\ref{fig:EPMstress} and~\ref{fig:EPMphaseDiagrams}.

In Figs. 2 and 3 of~\cite{PRL_SI}, we explore yielding in the soft
glassy rheology model in two different athermal protocols,
demonstrating the same scenario as in Fig.~\ref{fig:EPMstress} above
for this simpler elastoplastic model. In particular, in athermal
systems we always find ``brittle'' yielding as $\gdotbar\to 0$,
however small the stress overshoot.

In this work, we have studied shear-induced yielding of amorphous
materials as a function of initial sample annealing prior to shear. In
thermal systems we have demonstrated a gradual progression, with
increasing levels of annealing, from smoothly ``ductile'' yielding, in
which the sample remains homogeneous, to abruptly ``brittle''
yielding, in which it becomes strongly shear banded. We have shown
that this progression arises from an increase with annealing in the
size of an overshoot in the underlying stress-strain curve for
homogeneous shear. This in turn causes a shear banding instability
that becomes more severe with increasing annealing.  ``Ductile'' and
``brittle'' yielding thereby emerge as two limiting cases of a
continuum of yielding transitions, from gradual to catastrophic. In
contrast, we have shown that athermal systems with a stress overshoot
always show brittle yielding at low shear rates, however small the
overshoot. It remains to be understood how the thermal scenario crosses
over to the athermal one at low temperatures.

The scenarios put forward here differ notably from that recently
proposed in mean
field~\cite{ISI:000436245000061,popovic2018elastoplastic}, in which
``ductile'' and ``brittle'' yielding are separated by a critical
point, at which the underlying stress-strain curve for homogeneous
shear undergoes a qualitative change in shape as the degree of initial
sample annealing increases, as sketched in Fig.~\ref{fig:sketch}a).  To
discriminate between the scenario proposed here and that in
Ref.~\cite{ISI:000436245000061,popovic2018elastoplastic}, it would be
interesting to perform particle based simulations that disallow shear
banding, to access the shape of the underlying stress-strain curve for
homogeneous shear, either by using an algorithm that enforces a
homogeneous shear, or simulating a relatively small number of
particles, such that no $k-$mode can become unstable to shear
banding. Indeed, Fig. 2D of Ref.~\cite{ISI:000436245000061}, reporting
simulations for small numbers of particles, gives an indication that
the homogeneous stress-strain curve has the form sketched in
Fig.~\ref{fig:sketch}b,c) above. It further suggests that the smooth
athermal yielding seen for modest stress overshoots
in~\cite{ISI:000436245000061} may well stem from a finite size effect.

We thank Kirsten Martens, Romain Mari and Matthieu Wyart for
discussions; and the SOFI CDT, Durham University, EPSRC (EP/L015536/1)
and Schlumberger Cambridge Research for funding.

\bibliographystyle{apsrev2}

%

\newpage

\title{Supplemental material: ``Ductile and brittle yielding in thermal and athermal amorphous materials''}
\author{Hugh J. Barlow, James O. Cochran and Suzanne M. Fielding}
\affiliation{Department of Physics, Durham University, Science Laboratories,
  South Road, Durham DH1 3LE, UK}

{\large \bf Supplemental material: ``Ductile and brittle yielding in thermal and athermal amorphous materials''}

\vspace{0.35cm}


In this Supplemental Material, we present additional evidence
supporting our numerical observations in the main text: that slowly
sheared thermal systems show a smooth crossover from ``ductile'' to
``brittle'' yielding with increasing height of stress overshoot, and
therefore with increasing initial sample annealing prior to shear. In
contrast, slowly sheared athermal systems show ``brittle'' yielding
for any height of stress overshoot, however small.

We start in Sec.~\ref{sec:LSA} by performing a linear stability
analysis for the initial onset of shear banding as a system starts to
yield, in order to explain our numerical results (separately) for the
thermal fluidity and athermal elastoplastic models in the main
text. We show that the severity of shear banding (and so of yielding)
scales in the limit of slow shear, $\gdotbar\to 0$, as
$\exp(\Delta\Sigma)$ in the thermal model, with $\Delta\Sigma$ the
magnitude of stress drop from the stress overshoot. Accordingly, in
thermal systems a large enough stress overshoot is needed to see
severe banding and ``brittle'' failure. In contrast, in the athermal
model the degree of shear strain banding diverges (at the level of our
linear calculation) as $1/\partial\gamma\Sigma$ on approach to the
stress overshoot, $\partial\gamma\Sigma=0$. Accordingly, slowly
sheared athermal systems are predicted show ``brittle'' yielding,
however small the stress overshoot.

In Sec.~\ref{sec:SGR} we present additional numerical results
supporting those the main text, now in the context of the widely
studied soft glassy rheology (SGR) model~\cite{ISI:A1997WM06400048},
extended to include explicit spatial stress propagation between
elastoplastic elements, and with its noise-temperature parameter,
originally intended as a mean field description of this stress
propagation, here interpreted as the true thermal energy $k_{\rm
B}T$. In its thermal regime, we show that this gives the same
crossover from ``ductile'' to ``brittle'' yielding with increasing
height of stress overshoot, as for the thermal fluidity model in the
main text. In contrast, in its athermal regime it gives the same
scenario as the athermal elastoplastic model in the main text, with
``brittle'' yielding for any size of stress overshoot, however small,
in the limit of slow shear, $\gdotbar\to 0$.

Finally in Sec.~\ref{sec:delta}, we consider the effects of noise level on when
yielding first sets in.

\section{Linear stability analysis for onset of shear banding}
\label{sec:LSA}

\subsection{Thermal fluidity model}

To understand the numerical results of Figs.~2 and 3 in the main text
for the thermal fluidity model, and to explore the degree to which
they might generalise across fluidity models more widely, we cast that
model into a more general form, writing
\bea
\Sigma(t)&=&\sigma(y,t)+\eta\gdot(y,t),\nonumber\\
\dot{\sigma}(y,t)&=&f(\gdot,\sigma,\tau),\nonumber\\
\dot{\tau}(y,t)&=&g(\gdot,\sigma,\tau).
\eea
The specific fluidity model studied in our numerics in the main text
has
\bea
f&=&G\gdot-\sigma/\tau,\nonumber\\
g&=&1-|\gdot|\tau/(1+|\gdot|\tau_0).
\eea
(plus diffusive terms, which are relatively unimportant in the early
stages of any shear banding instability).

To consider the severity with which shear bands will form during any
deformation experiment, we write the system's state as the sum of a
homogeneous part plus an (initially) small heterogeneous perturbation:
\bea
\gdot(y,t)&=&\gdotbar+\delta\gdot(t)\exp(iky),\nonumber\\
\sigma(y,t)&=&\bar{\sigma}(t)+\delta\sigma(t)\exp(iky),\nonumber\\
\tau(y,t)&=&\bar{\tau}(t)+\delta\tau(t)\exp(iky).
\eea
The homogeneous part represents an underlying time-dependent base
state, as would pertain in a theoretically idealised deformation in
which shear bands are not allowed to form. The heterogeneous part is
the precursor to any shear bands. Substituting these into the model
equations and expanding to first order in the amplitude of the
perturbations, we find that the shear rate heterogeneity evolves as
\be
d\log(\delta\gdot)/dt=\omega+d\log(-f_{\tau}/f_{\gdot})/dt.
\ee
(This expression obtains in the limit of small solvent viscosities
$\eta\ll G\tau_0$, small diffusive terms $kl_0\ll 1$ and slow shear
rates $\gdotbar\tau_0\ll 1$, which is the physical regime of
interest.)  The first term, 
\be
\omega=-g_{\gdot}f_{\tau}/f_{\gdot}+g_{\tau},
\ee
is an eigenvalue which, when positive, indicates an instability to the
growth of heterogeneous shear bands. The second term represents a
time-dependent spinning of the eigenvector
$(\delta\tau,\delta\gdot)$. We find numerically that the first term
dominates the second after stress overshoot, which is the main regime
of interest in these thermal systems. The second term however gives
some modest growth of shear banding, even before the overshoot.

For any model of Maxwell-like form, $f=\gdot-\sigma/\tau$, and with
$\tau$ dynamics such that $g_{\gdot}=-\tau$, $g_{\tau}=-\gdot$ (which
is true for the fluidity model in the main text), it can be easily
shown that the eigenvalue
\be
\omega=-\partial_t \Sigma.
\ee
At the level of this linear calculation, therefore, the degree shear
banding $\delta\gdot(t)/\gdotbar$ after any stress drop of magnitude
$\Delta\Sigma$ from the stress overshoot scales as
\be
\frac{\delta\gdot(t)}{\gdotbar}=\frac{\delta\gdot_0}{\gdotbar}\exp(\Delta\Sigma),
\ee
to within small corrections, where $\delta\gdot_0/\gdotbar$ is the
small background shear rate heterogeneity due to noise. Systems with a
small stress overshoot are accordingly predicted to show ductile
yielding, and those with a large stress overshoot brittle
failure. Because the size of overshoot scales in the fluidity model as
an increasing function of $\gdotbar\tw$ for low $\gdotbar$ (contour
lines in Fig.~3 of the main text), deformation experiments with a high
$\gdotbar\tw$ show brittle failure and those with low $\gdotbar\tw$
show ductile yielding for all values of $\gdotbar$. This is indeed
confirmed numerically in Figs. 2c,d) of the main text.

\subsection{Athermal elastoplastic model}

The athermal elastoplastic model studied numerically in the main text
considers an ensemble of elastoplastic elements on each
streamline. The probability distribution $P(l,y,t)$ of local strains
$l$ among these elements on a streamline located at flow gradient
position $y$ evolves in time $t$ according to:
\be
\dot{P}(l,y,t)+\gdot(y,t)\, \partial_l P=-r(l)P+Y(y,t)\delta(l).
\ee
The local yielding rate $r(l)=\tau_0^{-1}$ when a local energy barrier
$E$ is exceeded, $\tfrac{1}{2}Gl^2>E$, and $r(l)=0$ otherwise, with
$G=1, E=1$ in our units. Accordingly the threshold yield strain for
local yielding $l_{\rm c}=\sqrt{2}$. The elastoplastic stress on any
streamline is the average of the elemental ones: $\sigma(y, t)=G\int dl\,
l P(l, y, t)$. As usual we impose force balance by insisting that the
total stress $\Sigma$ is uniform across streamlines:
\be
\Sigma(t)=\sigma(y,t)+\eta\gdot(y,t).
\ee

In the limit of slow shear, $\gdot\to 0$, elements yield essentially
instantaneously when they reach the threshold strain $l_{\rm c}$. The average
elastoplastic stress on any streamline accordingly evolves in this
limit  according to the
simpler equation:
\be
\dot{\sigma}(y,t)=\gdot(y,t)\left[1-l_{\rm c}P_{\rm c}(y,t)\right],
\ee
in which $P_{\rm c}(y,t)$ is the probability of an element on a
streamline at $y$ having a local strain just below the threshold
$l_{\rm c}$ at time $t$. (We assume a positive shear rate $\gdot>0$.) 
Assuming that no element has approached this threshold more than once
since shearing started (which is true up to the stress overshoot for
most parameter regimes considered), it is trivial to show that $P_{\rm
c}(y,t)=P_0(l_{\rm c}-\gamma(y,t))$, where $P_0(l)$ is the initial
distribution of local strains, giving
\be
\dot{\sigma}(y,t)=\gdot(y,t)\left[1-l_{\rm c}P_0(l_{\rm c}-\gamma(y,t))\right].
\ee

To consider the severity with which shear bands will form during any
deformation experiment, we write the system's state as the sum of a
homogeneous part plus an (initially) small heterogeneous perturbation:
\bea
\gamma(y,t)&=&\bar{\gamma}(t)+\delta\gamma(t)\exp(iky),\nonumber\\
\gdot(y,t)&=&\gdotbar+\delta\gdot(t)\exp(iky),\nonumber\\
\sigma(y,t)&=&\bar{\sigma}(t)+\delta\sigma(t)\exp(iky).\nonumber\\
\eea
The homogeneous part represents an underlying time-dependent base
state, as would pertain in a theoretically idealised deformation in
which shear bands are not allowed to form. 
The heterogeneous part is the precursor to any shear
bands. Substituting these into the model equations and expanding to
first order in the amplitude of the perturbations, we find in the
limit $\eta\to 0$ that the shear strain heterogeneity after any
average imposed deformation to the sample as a whole, $\bar{\gamma}$,
obeys:
\be
\delta\gamma(\bar{\gamma})=\frac{\delta\gamma(0)}{\partial_\gamma\Sigma}.
\ee
It is accordingly predicted to diverge on approach to the stress
overshoot, $\partial_\gamma\Sigma=0$: slowly sheared athermal
materials are thus predicted to show ``brittle'' yielding, however
small the stress overshoot.

\section{Soft glassy rheology (SGR) model}
\label{sec:SGR}

The soft glassy rheology (SGR) model~\cite{ISI:A1997WM06400048}
considers an ensemble of elastoplastic elements, each of which
corresponds to a local mesoscopic region of material.  Given a shear
rate $\gdot$, each element builds up a local elastic shear strain $l$
according to $\dot{l}=\gdot$, giving a shear stress $Gl$, where $G$ is
a constant modulus. This stress is intermittently released by local
plastic yielding events, each modelled as hopping of an element over a
strain-modulated energy barrier $E$, governed by a temperature
parameter $x$, with a stochastic yielding rate
$r(E,l)=\tau_0^{-1}\rm{min}\left\{1,\exp[-(E-\tfrac{1}{2}kl^2)/x]\right\}$,
in which $\tau_0$ is a microscopic attempt
time~\cite{ISI:A1997WM06400048}. Upon yielding, any element resets its
local stress to zero and chooses its new energy barrier from an
exponential distribution $\rho(E)=\exp(-E/x_{\rm g})/x_{\rm g}$, which
has a characteristic $\langle E \rangle \sim x_{\rm g}$.  This results
in a broad spectrum of yielding times, $P(\tau)$, and a glass phase
for $x<x_{\rm g}$, in which a sample shows ageing before deformation
commences.

The probability distribution $P(E,l,t)$ of local strains evolves according to:
\be
\dot{P}(E,l,t)+\gdot\, \partial_l P=-r(E,l)P+Y(t)\rho(E)\delta(l).
\ee
Here $Y(t)=\int dl\,dE\, r(E,l)\,P(E,l,t)$ is the ensemble average local
yielding rate and  $\delta(l)$ is the Dirac delta function.  The total
elastoplastic stress $\sigma(t)=G \int dl\, \int dE\, l\, P(E, l,t)$.

So far, we have assumed homogeneous flow in the SGR model, without
accounting for spatial stress propagation following any local yielding
event. To account for non-uniform shear deformations, we now take
$n=1...N$ streamlines at discretised flow-gradient positions
$y=0...L_y$, with periodic boundary conditions. The distribution
$P(E,l,y,t)$ on any streamline then obeys:
\be
\dot{P}(E,l,y,t)+\gdot(y,t)\, \partial_l P=-r(E,l)P+Y(y,t)\rho(E)\delta(l),
\ee
with streamline yielding rate $Y(y,t)=\int dl\, \int dE\,
r(E,l)\,P(E,l,y,t)$ and elastoplasic stress $\sigma(y,t)=G \int dl\, dE\,
l\, P(E,l,y, t)$.  Given an imposed average shear rate $\gdotbar$ across
the sample as a whole, the shear rate on each streamline is calculated
by enforcing force balance:
$\sigma(y,t)+\eta\gdot(y,t)=\bar\sigma(t)+\eta\gdotbar$, with
$\bar\sigma(t)=\tfrac{1}{L_y}\int dy \sigma(y,t)$.  This ensures 1D
stress propagation on a timescale $\eta/G$ following any local
yielding event, recovering the 1D projection of the Eshelby propagator
of 2D lattice elastoplastic models~\cite{nicolas2018deformation}.

We simulate this model by evolving $M=80000$ SGR elements on each of $N=20$
streamlines, with force balance across streamlines as described.
Adjacent streamlines are further weakly coupled by adjusting the
stress of three randomly chosen elements on each adjacent streamline
an amount $wl(-1,+2,-1)$ following any yielding event of size $l$,
with $w=0.05$ a small parameter. This mimics the stress diffusion term the
continuum Eqn.~2 of the main text.  We rescale strain, stress, time
and length so that $x_{\rm g}=G=\tau_0=L_y=1$.  The solvent viscosity
$\eta\ll G\tau_0=1$ is unimportant to the physics we describe. We use
typical values between $\eta=0.01$ and $0.05$, but find no changes to
our results on reducing $\eta$ further.

It is worth pausing to compare the SGR model studied in this
Supplemental Material with the athermal elastoplastic model in the
main text. A key difference between the two models is that in the SGR
model an elastoplastic element can be activated into yielding even
before the top of its energy barrier is reached, by virtue of the
temperature parameter, $x$. In contrast, the elastoplastic model in
the main text is athermal. (It sets $x=0$ upfront.) In some
interpretations of the SGR model, $x$ is taken as a mechanical noise
temperature. Here we take it to be the true thermal temperature. The
other difference between the two models is that the SGR model has a
prior distribution of trap depths $\rho(E)\sim\exp(-E/\xg)$, whereas
$E=1$ for all elements in the elastoplastic model in the main
text. The combination of the exponential prior with the thermal
activation factor captures ageing prior to deformation in the SGR
model. In the elastoplastic model of the main text, in contrast, the
initial condition at the start of shear must be inserted
`artificially' because that model shows no dynamical evolution before
deformation commences. The two models are otherwise identical.

We now present numerical results for the predictions of the SGR model
for yielding in three different commonly studied annealing protocols.
The first protocol is thermal, the second and third are athermal. In
each protocol, we seed the formation of shear bands by slightly
perturbing the trap depths, after annealing and immediately prior to
shear, as $E\to E\left[1+\delta\sin(2\pi y/L_y)\right]$ with
$\delta=10^{-3}$.

\subsection{Results: thermal SGR model after ageing}

We consider first a sample freshly prepared in a rejuvenated state at
time $t=-\tw$ then held at a constant temperature in the glass phase
$0<x<\xg=1$ for a waiting (annealing) time $\tw$. It is then sheared
at a constant rate $\gdotbar$ for all times $t>0$, with $x$ still held
constant.  In this protocol, better annealed samples correspond to
longer ageing times $\tw$. It is the same protocol as studied for the
thermal fluidity model in the main text, although temperature does not
appear as an explicit parameter in that model.

\begin{figure}[!t]
\includegraphics[width=8.5cm]{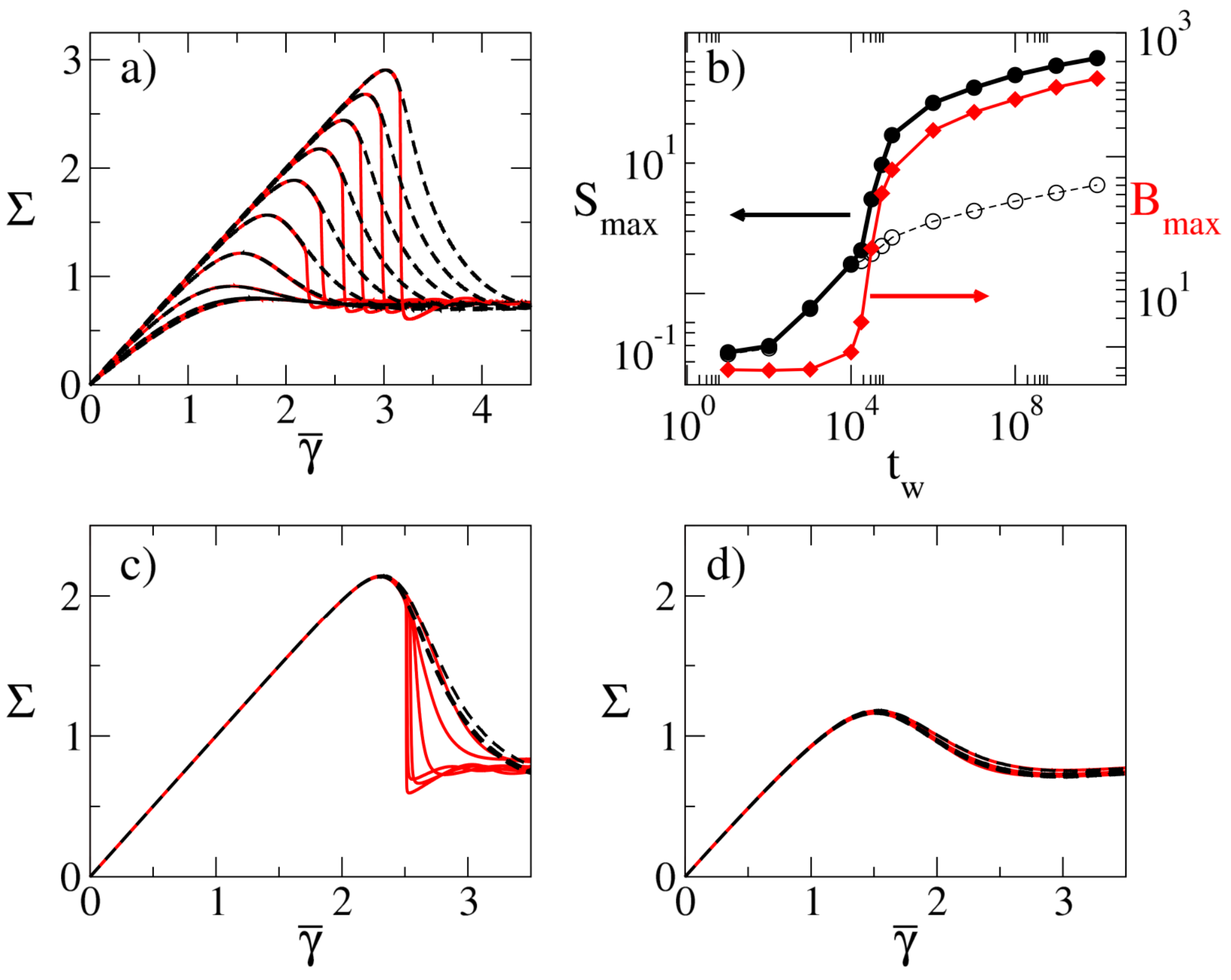}
\caption{{\bf a)} Stress versus strain in the  SGR model with homogeneous flow enforced (dashed lines) and shear banding allowed (solid lines) for waiting times $t_w = 10^1 , 10^2\cdots 10^{10}$ in curves left to right. Imposed shear rate $\gdotbar=\sqrt{2}\times 10^{-3}$. {\bf b)} (left vertical axis)  steepest negative slope in stress versus strain curve with homogeneous flow enforced (open symbols) and allowing banding (closed black symbols) as function of sample age $\tw$ for a fixed $\gdotbar=\sqrt{2}\times 10^{-3}$; (right vertical axis) corresponding maximum degree of shear banding during deformation. $\Bmax$ and $\Smax$ are both  averaged over $50$  simulations at each $\tw$. {\bf c)} and {\bf d)} show curves in same format as {\bf a)}, but now for a fixed $\gdotbar \tw=10^4$ giving a large stress overshoot (c) and a fixed $\gdotbar \tw=10$ giving a small stress overshoot (d), for imposed  shear rates $\gdotbar=10^{-n}$ with $n=1.0,~1.5,~2.0,\cdots 4.0$ in solid curves right to left in c). Temperature $x = 0.3$.}
\label{fig:SGR_Moorcroft}
\end{figure}

Fig.~\ref{fig:SGR_Moorcroft}a) shows as dashed lines the underlying
stress-strain curves, computed within the SGR model in this thermal
protocol, for deformations in which the shear field is constrained to
remain homogeneous. Results are shown for several different sample
ages $\tw$ at a fixed value of the imposed shear rate. As in Fig. 2a)
of the main text for the thermal fluidity model, each curve shows a
stress overshoot, the size of which increases with increasing degree
of sample annealing before deformation commences (increasing $\tw$).

The solid lines in Fig.~\ref{fig:SGR_Moorcroft}a) show the results of
separate calculations in which shear bands are allowed to form. For
poorly annealed samples, the stress-strain curve closely follows that
of the homogeneous calculation, indicating that the deformation
remains homogeneous, to good approximation, with no shear bands
forming. Yielding accordingly remains a ductile process. For better
annealed samples, the stress drops precipitously below that computed
when homogeneity is enforced, leading to abrupt yielding. The maximal
gradient of stress drop and maximal degree of shear banding in any
deformation experiment are plotted as a function of sample age in
Fig.~\ref{fig:SGR_Moorcroft}b). (We define the degree of shear banding
as the standard deviation in $\gdot$ across $y$, normalised by the
imposed shear rate $\gdotbar$.) This shows the same behaviour as for
the thermal fluidity model in Fig.~2b) of the main text: the severity
of the stress drop and shear banding increase with increasing
annealing, $\tw$, marking a crossover from ``ductile'' to ``brittle''
yielding as the height of the stress overshoot increases.

\begin{figure}[!t]
\includegraphics[width=8.5cm]{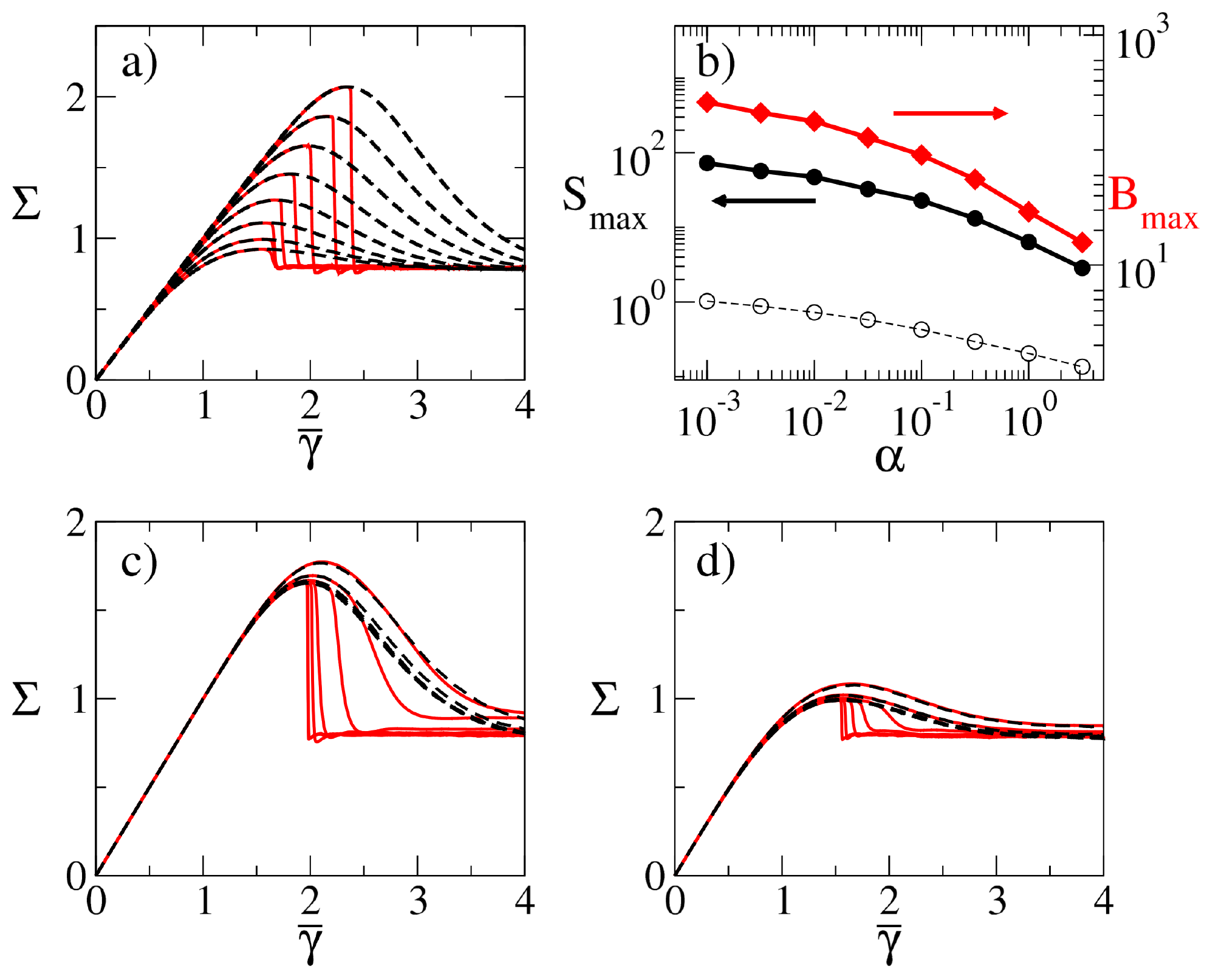}
\caption{{\bf a)} Stress versus strain in the SGR model with homogeneous flow enforced (dashed lines) and shear banding allowed (solid lines) for cooling rates $\alpha = 10^{-n}$ with $n = 0.5,~1.0,~1.5\cdots 3.0$ in curves left to right. Imposed shear rate $\gdotbar=\sqrt{2}\times 10^{-3}$. {\bf b)} (left vertical axis) steepest negative slope in stress versus strain curve with homogeneous flow enforced (open symbols) and allowing banding (closed black symbols) as a function of cooling rate $\alpha$ for a fixed $\gdotbar=\sqrt{2}\times 10^{-3}$; (right vertical axis) corresponding maximum degree of shear banding during deformation. $\Bmax$ and $\Smax$ are averaged over $50$ simulations at each $\alpha$. {\bf c)} and {\bf d)} show curves in same format as {\bf a)}, but now for a  fixed $\alpha = 10^{-2}$ giving a large stress overshoot (c) and  for a fixed $\alpha = 10^{0}$ giving a small stress overshoot (d), for imposed shear rates  $\gdotbar=\sqrt{2} \times 10^{-n}$ with $n=1.0,~1.5,~2.0,\cdots 4.0$ in  curves right to left. Temperature $x = 0.0$.}
\label{fig:SGR_DelGado}
\end{figure}
\begin{figure}[!t]
\includegraphics[width=8.5cm]{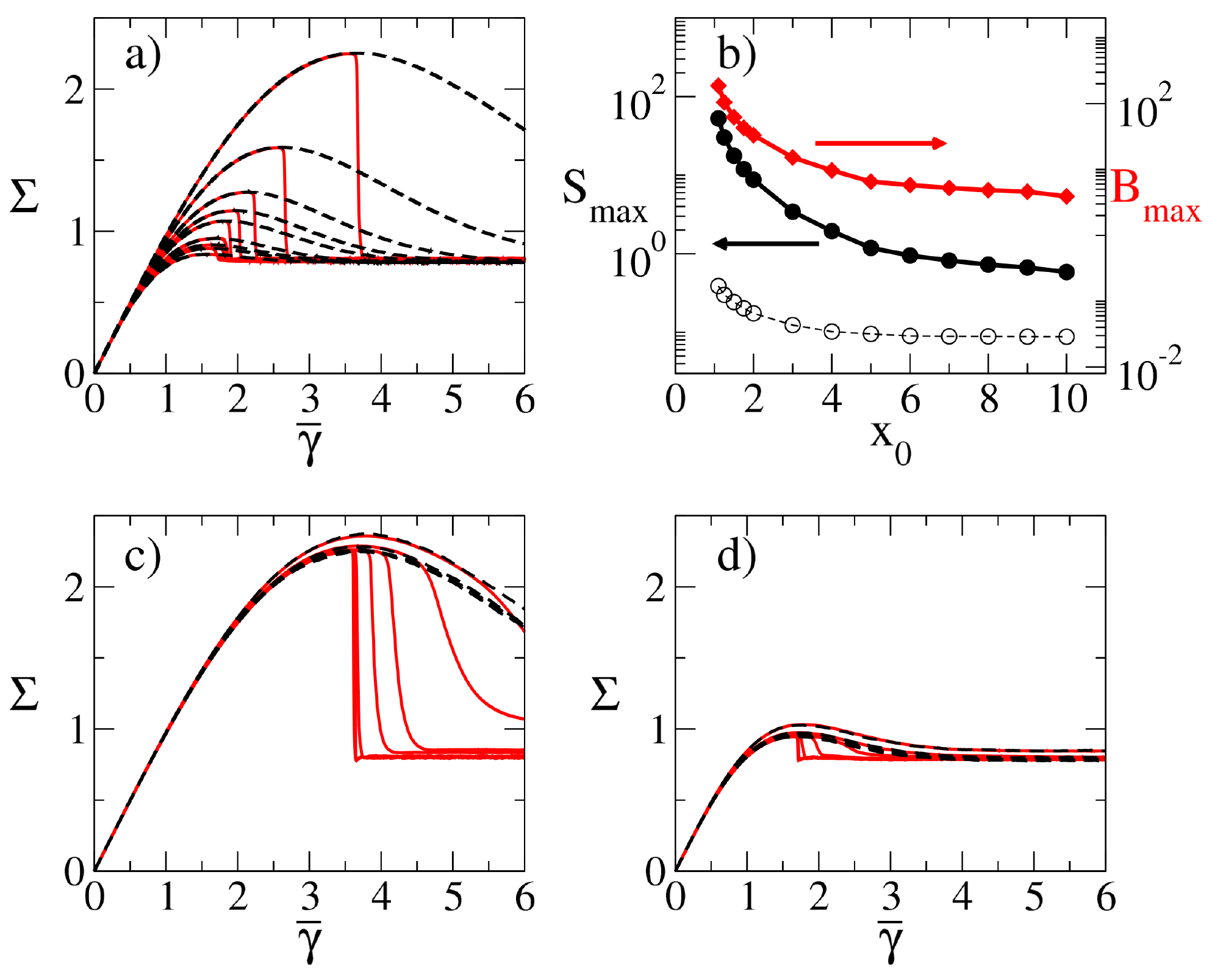}
\caption{{\bf a)} Stress versus strain in the SGR model with homogeneous flow enforced (dashed lines) and shear banding allowed (solid lines) for annealing temperatures $x_0 = 1.1$, $1.25$, $1.5$, $1.75$, $2.0$, $3.0$, $4.0$, $5.0$, $6.0$, $7.0$, $8.0$, $9.0$, $10.0$ in curves right to left. Imposed shear rate $\gdotbar=\sqrt{2}\times 10^{-3}$. {\bf b)} (left vertical axis) steepest negative slope in stress versus strain curve with homogeneous flow enforced (open circles) and allowing banding (closed black circles) as a function of annealing temperature, for a fixed $\gdotbar=\sqrt{2}\times 10^{-3}$; (right vertical axis) corresponding maximum degree of shear banding during deformation. $\Bmax$ and $\Smax$ are averaging over $50$  simulations at each $x_0$. {\bf c)} and {\bf d)} show curves in same format as {\bf a)}, but now for a  fixed $x_0 = 1.1$ giving a large stress overshoot (c) and  for a fixed $x_0=3.0$ giving a small stress overshoot (d), for imposed shear rates $\gdotbar=\sqrt{2} \times 10^{-n}$ with $n=1.0,~1.5,~2.0,\cdots 4.0$  in curves right to left. Temperature $x = 0.0$.}
\label{fig:SGR_Berthier}
\end{figure}

As in the thermal fluidity model, the height of the stress overshoot
scales as an increasing function of $\gdotbar\tw$. In
Fig.~\ref{fig:SGR_Moorcroft}c) we show results for a fixed large value
of $\gdotbar\tw$, giving a stress overshoot of fixed large height, for
several different values of $\gdotbar$. For high $\gdotbar$, we see
``ductile'' yielding, with ``brittle'' yielding setting in as
$\gdotbar\to 0$ (for this fixed large $\gdotbar\tw$).  In
Fig.~\ref{fig:SGR_Moorcroft}d), we show results for a fixed smaller
value of $\gdotbar\tw$, giving a smaller stress overshoot. Here we see
``ductile'' yielding even at very low values of the imposed strain
rate. The same scenario was seen in Figs. 2c,d) of the main text for
the thermal fluidity model.

To summarise, the SGR model in its thermal regime shows the same
scenario as the thermal fluidity model in the main text. This can be
seen by the striking similarity between each of the panels of
Fig.~\ref{fig:SGR_Moorcroft}a-d) for the thermal SGR model with its
counterpart in Fig. 2 of the main text for the thermal fluidity model.
In particular, in the limit of slow shear, $\gdotbar\to 0$, the
severity of shear banding and therefore of yielding increase with
increasing height of the stress overshoot, which is in turn set by
$\gdotbar\tw$.

\subsection{Results: athermal SGR model after slow cooling.}

We now consider an annealing protocol that consists of equilibrating
the sample to a high initial temperature $x=x_0=5.0>\xg$ then cooling
to zero temperature at a constant cooling rate $\alpha$, such that
$x(t)=x_0-\alpha t$, before shearing at a constant rate $\gdotbar$ for
all subsequent times, at $x=0$.  In this case, a slower cooling rate
$\alpha$ corresponds to a better annealed sample.

Fig.~\ref{fig:SGR_DelGado}a) shows as dashed lines the stress
$\Sigma(\gammabar)$ as a function of the accumulating strain
$\gammabar=\gdotbar t$, calculated by imposing that the shear must
remain homogeneous across the sample, for several levels of annealing
prior to shear, at a single imposed shear rate $\gdotbar$. Each curve
shows an initial solid-like elastic regime in which the stress
increases linearly with strain, before the stress declines as plastic
yielding sets in.  As can be seen, more strongly annealed samples
(lower $\alpha$) show a larger initial elastic regime.

We then performed separate calculations in which shear bands can
form. The resulting stress-strain curves are shown by solid lines in
Fig.~\ref{fig:SGR_DelGado}a). Once the stress overshoot is reached,
the stress in these heterogeneous calculations falls precipitously
below that of the homogeneous calculations: shear bands form, leading
to ``brittle'' failure. By comparing each of
Figs.~\ref{fig:SGR_DelGado}a,c,d) with its counterpart subpanel in
Fig.~~\ref{fig:SGR_Moorcroft}a,c,d), we see an important difference
between the thermal and athermal regimes of the SGR model. In the
thermal regime, a large enough stress overshoot is required to see
``brittle'' failure. In contrast, in its athermal regime it shows
``brittle'' failure in slow shear, $\gdotbar\to 0$, however small the overshoot.

\subsection{Results: athermal SGR model after rapid quenching.}

\begin{figure}[!t]
\includegraphics[width=8.0cm]{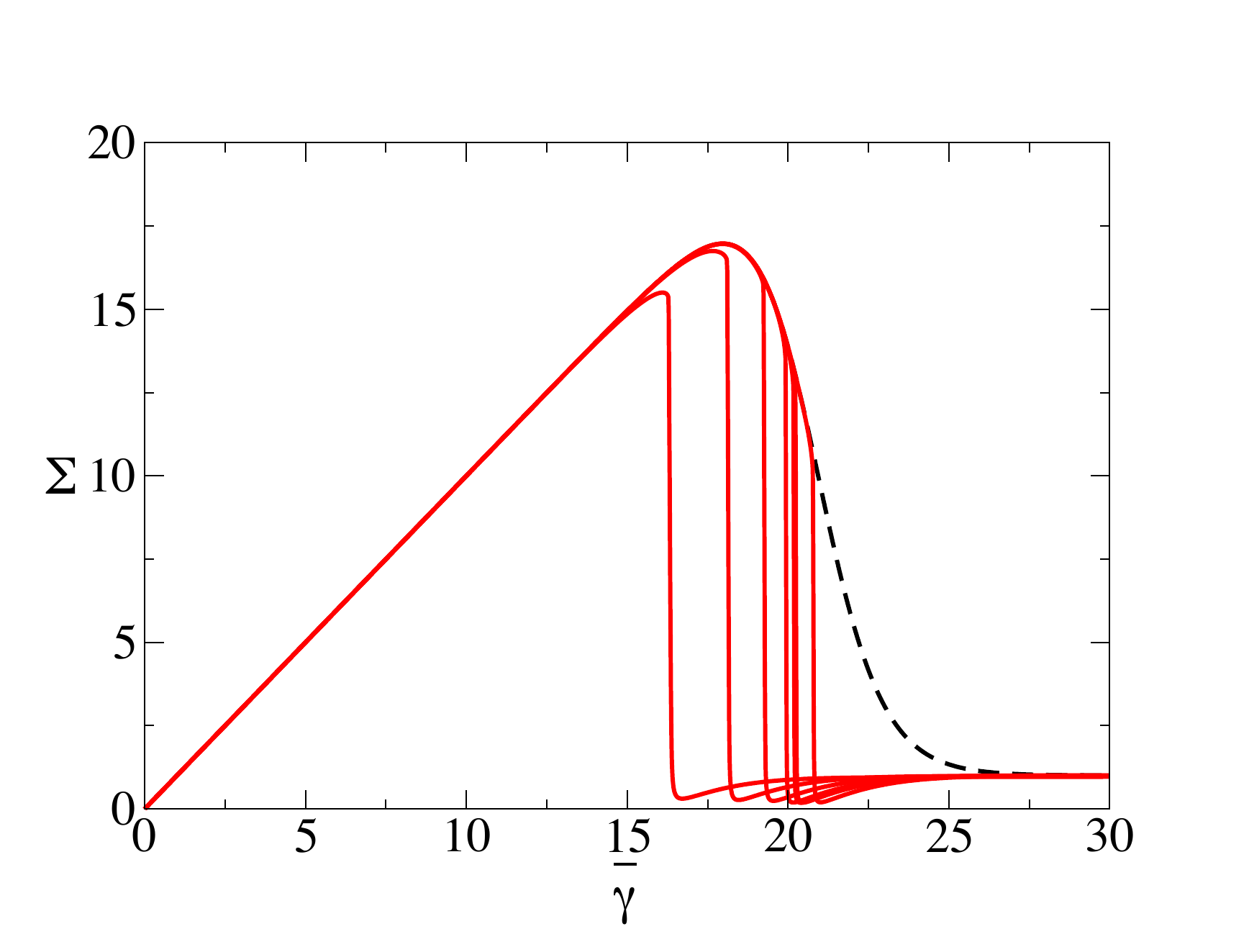}
\caption{Stress as a function of strain in the thermal fluidity model of the main text, with homogeneous flow enforced (dashed lines) and shear banding allowed (solid lines), for an imposed shear rate $\gdotbar=10^{-3}$ and  waiting time $\tw=10^{12}$. Noise levels $\delta=10^{-n}$ with $n=3.0, 3.167, 3.33, 3.5, 4.0, 4.5, 5.0$  in curves left to right.  Interface width $l=10^{-3}$. Numerical timestep $Dt=0.01$, meshsize $Dy=1/3000$.}
\label{fig:delta}
\end{figure}

We finally consider an annealing protocol that consists of
equilibrating the sample to an initial temperature $x_0>\xg$, then at
some time $t=0$ suddenly jumping the temperature to $x=0$, before
shearing at a constant rate $\gdotbar$ for all subsequent times, at
$x=0$. In this protocol, a lower initial equilibration temperature
$x_0$ corresponds to a better annealed sample.

Fig.~\ref{fig:SGR_Berthier}a) shows as dashed lines the stress
$\Sigma(\gammabar)$ as a function of the accumulating strain
$\gammabar=\gdotbar t$, calculated by imposing that the shear must
remain homogeneous across the sample, for several levels of annealing
prior to shear, at a single imposed shear rate $\gdotbar$. Each curve
shows an initial solid-like elastic regime in which the stress
increases linearly with strain, before the stress declines as plastic
yielding sets in.  As can be seen, more strongly annealed samples
(lower $x_0$) show a larger initial elastic regime.

We then performed separate calculations in which shear bands can
form. The resulting stress-strain curves are shown by solid lines in
Fig.~\ref{fig:SGR_Berthier}a). Once the stress overshoot is reached,
the stress in these heterogeneous calculations falls precipitously
below that of the homogeneous calculations: shear bands form, leading
to ``brittle'' failure. By comparing each of
Figs.~\ref{fig:SGR_Berthier}a,c,d) with its counterpart subpanel in
Fig.~\ref{fig:SGR_Moorcroft}, we once again see an important difference
between the thermal and athermal regimes of the SGR model. In the
thermal regime, a large enough stress overshoot is required to see
``brittle'' failure. In contrast, in its athermal regime it shows
``brittle'' failure in the limit of slow shear, $\gdotbar\to 0$,
however small the overshoot.

{\em We have now demonstrated this key difference between yielding in
thermal and athermal systems in three ways:} (i) numerically by
comparing the thermal fluidity model with the athermal elastoplastic
model in the main text, (ii) analytically by performing linear
stability analyses in the thermal fluidity model and the athermal
elastoplastic model in Sec. 1 of this Supplemental Material and
(iii) numerically by comparing the SGR model's behaviour in its
thermal and athermal regimes.

\section{Effect of noise level on yielding onset}
\label{sec:delta}

In the main text, we argued that shear banding, and the associated
brittle failure to which it leads, arises from a linear instability of
a state of initially homogeneous deformation, in the regime where the
stress is a declining function of strain. Any linear instability must
of course be seeded by noise or initial disorder. We explore finally
the effect of noise level on the dynamics of shear banding and
yielding. Fig.~\ref{fig:delta} shows counterpart curves to those of
Fig. 2a) of the main text, for the particular value $\tw=10^{12}$, now
for different levels of the noise $\delta$. As can be seen, for
increasing $\delta$ the stress drop indicative of shear banding occurs
earlier and earlier. Indeed, for large enough $\delta$ it can occur
even before the regime of declining stress versus strain for an
underlying state of homogeneous deformation, and therefore before the
eigenvalue signifying instability to banding becomes positive in a
thermal system. The nucleation of a transition via a finite amplitude
perturbation before a regime of linear instability is well known in
the context of equilibrium phase transitions. It was discussed with
regards yielding of amorphous materials in
Ref.~\cite{popovic2018elastoplastic}.


\begin{thebibliography}{43}%
\makeatletter
\providecommand \@ifxundefined [1]{%
 \@ifx{#1\undefined}
}%
\providecommand \@ifnum [1]{%
 \ifnum #1\expandafter \@firstoftwo
 \else \expandafter \@secondoftwo
 \fi
}%
\providecommand \@ifx [1]{%
 \ifx #1\expandafter \@firstoftwo
 \else \expandafter \@secondoftwo
 \fi
}%
\providecommand \natexlab [1]{#1}%
\providecommand \enquote  [1]{``#1''}%
\providecommand \bibnamefont  [1]{#1}%
\providecommand \bibfnamefont [1]{#1}%
\providecommand \citenamefont [1]{#1}%
\providecommand \href@noop [0]{\@secondoftwo}%
\providecommand \href [0]{\begingroup \@sanitize@url \@href}%
\providecommand \@href[1]{\@@startlink{#1}\@@href}%
\providecommand \@@href[1]{\endgroup#1\@@endlink}%
\providecommand \@sanitize@url [0]{\catcode `\\12\catcode `\$12\catcode
  `\&12\catcode `\#12\catcode `\^12\catcode `\_12\catcode `\%12\relax}%
\providecommand \@@startlink[1]{}%
\providecommand \@@endlink[0]{}%
\providecommand \doibase [0]{http://dx.doi.org/}%
\providecommand \selectlanguage [0]{\@gobble}%
\providecommand \bibinfo  [0]{\@secondoftwo}%
\providecommand \bibfield  [0]{\@secondoftwo}%
\providecommand \translation [1]{[#1]}%
\providecommand \BibitemOpen [0]{}%
\providecommand \bibitemStop [0]{}%
\providecommand \bibitemNoStop [0]{.\EOS\space}%
\providecommand \EOS [0]{\spacefactor3000\relax}%
\providecommand \BibitemShut  [1]{\csname bibitem#1\endcsname}%
\let\auto@bib@innerbib\@empty
\bibitem [{\citenamefont {Bonn}\ \emph {et~al.}(2017)\citenamefont {Bonn},
  \citenamefont {Denn}, \citenamefont {Berthier}, \citenamefont {Divoux},\ and\
  \citenamefont {Manneville}}]{ISI:000407999000001}%
  \BibitemOpen
  \bibfield  {author} {\bibinfo {author} {\bibfnamefont {D.}~\bibnamefont
  {Bonn}}, \bibinfo {author} {\bibfnamefont {M.~M.}\ \bibnamefont {Denn}},
  \bibinfo {author} {\bibfnamefont {L.}~\bibnamefont {Berthier}}, \bibinfo
  {author} {\bibfnamefont {T.}~\bibnamefont {Divoux}}, \ and\ \bibinfo {author}
  {\bibfnamefont {S.}~\bibnamefont {Manneville}},\ }\href {\doibase
  {10.1103/RevModPhys.89.035005}} {\bibfield  {journal} {\bibinfo  {journal}
  {{Rev. Mod. Phys.}}\ }\textbf {\bibinfo {volume} {{89}}},\ \bibinfo {pages}
  {035005} (\bibinfo {year} {{2017}})}\BibitemShut {NoStop}%
\bibitem [{\citenamefont {Coussot}(2018)}]{ISI:000419993800001}%
  \BibitemOpen
  \bibfield  {author} {\bibinfo {author} {\bibfnamefont {P.}~\bibnamefont
  {Coussot}},\ }\href {\doibase {10.1007/s00397-017-1055-7}} {\bibfield
  {journal} {\bibinfo  {journal} {{Rheol. Acta}}\ }\textbf {\bibinfo {volume}
  {{57}}},\ \bibinfo {pages} {1} (\bibinfo {year} {{2018}})}\BibitemShut
  {NoStop}%
\bibitem [{\citenamefont {Bonn}\ and\ \citenamefont
  {Denn}(2009)}]{ISI:000266878700032}%
  \BibitemOpen
  \bibfield  {author} {\bibinfo {author} {\bibfnamefont {D.}~\bibnamefont
  {Bonn}}\ and\ \bibinfo {author} {\bibfnamefont {M.~M.}\ \bibnamefont
  {Denn}},\ }\href {\doibase {10.1126/science.1174217}} {\bibfield  {journal}
  {\bibinfo  {journal} {{Science}}\ }\textbf {\bibinfo {volume} {{324}}},\
  \bibinfo {pages} {1401} (\bibinfo {year} {{2009}})}\BibitemShut {NoStop}%
\bibitem [{\citenamefont {Coussot}(2007)}]{ISI:000246368400004}%
  \BibitemOpen
  \bibfield  {author} {\bibinfo {author} {\bibfnamefont {P.}~\bibnamefont
  {Coussot}},\ }\href {\doibase {10.1039/b611021p}} {\bibfield  {journal}
  {\bibinfo  {journal} {{Soft Matter}}\ }\textbf {\bibinfo {volume} {{3}}},\
  \bibinfo {pages} {528} (\bibinfo {year} {{2007}})}\BibitemShut {NoStop}%
\bibitem [{\citenamefont {Denn}\ and\ \citenamefont
  {Bonn}(2011)}]{ISI:000293295000002}%
  \BibitemOpen
  \bibfield  {author} {\bibinfo {author} {\bibfnamefont {M.~M.}\ \bibnamefont
  {Denn}}\ and\ \bibinfo {author} {\bibfnamefont {D.}~\bibnamefont {Bonn}},\
  }\href {\doibase {10.1007/s00397-010-0504-3}} {\bibfield  {journal} {\bibinfo
   {journal} {{Rheol. Acta}}\ }\textbf {\bibinfo {volume} {{50}}},\ \bibinfo
  {pages} {307} (\bibinfo {year} {{2011}})}\BibitemShut {NoStop}%
\bibitem [{\citenamefont {Hufnagel}\ \emph {et~al.}(2016)\citenamefont
  {Hufnagel}, \citenamefont {Schuh},\ and\ \citenamefont
  {Falk}}]{ISI:000374617600037}%
  \BibitemOpen
  \bibfield  {author} {\bibinfo {author} {\bibfnamefont {T.~C.}\ \bibnamefont
  {Hufnagel}}, \bibinfo {author} {\bibfnamefont {C.~A.}\ \bibnamefont {Schuh}},
  \ and\ \bibinfo {author} {\bibfnamefont {M.~L.}\ \bibnamefont {Falk}},\
  }\href {\doibase {10.1016/j.actamat.2016.01.049}} {\bibfield  {journal}
  {\bibinfo  {journal} {{Acta Mater.}}\ }\textbf {\bibinfo {volume} {{109}}},\
  \bibinfo {pages} {375} (\bibinfo {year} {{2016}})}\BibitemShut {NoStop}%
\bibitem [{\citenamefont {Greer}\ \emph {et~al.}(2013)\citenamefont {Greer},
  \citenamefont {Cheng},\ and\ \citenamefont {Ma}}]{ISI:000320906700001}%
  \BibitemOpen
  \bibfield  {author} {\bibinfo {author} {\bibfnamefont {A.~L.}\ \bibnamefont
  {Greer}}, \bibinfo {author} {\bibfnamefont {Y.~Q.}\ \bibnamefont {Cheng}}, \
  and\ \bibinfo {author} {\bibfnamefont {E.}~\bibnamefont {Ma}},\ }\href
  {\doibase {10.1016/j.mser.2013.04.001}} {\bibfield  {journal} {\bibinfo
  {journal} {{Mater. Sci. Eng. R. Rep.}}\ }\textbf {\bibinfo {volume} {{74}}},\
  \bibinfo {pages} {71} (\bibinfo {year} {{2013}})}\BibitemShut {NoStop}%
\bibitem [{\citenamefont {Divoux}\ \emph {et~al.}(2011)\citenamefont {Divoux},
  \citenamefont {Barentin},\ and\ \citenamefont
  {Manneville}}]{ISI:000295085700080}%
  \BibitemOpen
  \bibfield  {author} {\bibinfo {author} {\bibfnamefont {T.}~\bibnamefont
  {Divoux}}, \bibinfo {author} {\bibfnamefont {C.}~\bibnamefont {Barentin}}, \
  and\ \bibinfo {author} {\bibfnamefont {S.}~\bibnamefont {Manneville}},\
  }\href {\doibase {10.1039/c1sm05740e}} {\bibfield  {journal} {\bibinfo
  {journal} {{Soft Matter}}\ }\textbf {\bibinfo {volume} {{7}}},\ \bibinfo
  {pages} {9335} (\bibinfo {year} {{2011}})}\BibitemShut {NoStop}%
\bibitem [{\citenamefont {Divoux}\ \emph {et~al.}(2010)\citenamefont {Divoux},
  \citenamefont {Tamarii}, \citenamefont {Barentin},\ and\ \citenamefont
  {Manneville}}]{ISI:000277945900061}%
  \BibitemOpen
  \bibfield  {author} {\bibinfo {author} {\bibfnamefont {T.}~\bibnamefont
  {Divoux}}, \bibinfo {author} {\bibfnamefont {D.}~\bibnamefont {Tamarii}},
  \bibinfo {author} {\bibfnamefont {C.}~\bibnamefont {Barentin}}, \ and\
  \bibinfo {author} {\bibfnamefont {S.}~\bibnamefont {Manneville}},\ }\href
  {\doibase {10.1103/PhysRevLett.104.208301}} {\bibfield  {journal} {\bibinfo
  {journal} {{Phys. Rev. Lett.}}\ }\textbf {\bibinfo {volume} {{104}}},\
  \bibinfo {pages} {208301} (\bibinfo {year} {{2010}})}\BibitemShut {NoStop}%
\bibitem [{\citenamefont {Divoux}\ \emph {et~al.}(2012)\citenamefont {Divoux},
  \citenamefont {Tamarii}, \citenamefont {Barentin}, \citenamefont {Teitel},\
  and\ \citenamefont {Manneville}}]{ISI:000301801100015}%
  \BibitemOpen
  \bibfield  {author} {\bibinfo {author} {\bibfnamefont {T.}~\bibnamefont
  {Divoux}}, \bibinfo {author} {\bibfnamefont {D.}~\bibnamefont {Tamarii}},
  \bibinfo {author} {\bibfnamefont {C.}~\bibnamefont {Barentin}}, \bibinfo
  {author} {\bibfnamefont {S.}~\bibnamefont {Teitel}}, \ and\ \bibinfo {author}
  {\bibfnamefont {S.}~\bibnamefont {Manneville}},\ }\href {\doibase
  {10.1039/c2sm06918k}} {\bibfield  {journal} {\bibinfo  {journal} {{Soft
  Matter}}\ }\textbf {\bibinfo {volume} {{8}}},\ \bibinfo {pages} {4151}
  (\bibinfo {year} {{2012}})}\BibitemShut {NoStop}%
\bibitem [{\citenamefont {Grenard}\ \emph {et~al.}(2014)\citenamefont
  {Grenard}, \citenamefont {Divoux}, \citenamefont {Taberlet},\ and\
  \citenamefont {Manneville}}]{ISI:000332461800012}%
  \BibitemOpen
  \bibfield  {author} {\bibinfo {author} {\bibfnamefont {V.}~\bibnamefont
  {Grenard}}, \bibinfo {author} {\bibfnamefont {T.}~\bibnamefont {Divoux}},
  \bibinfo {author} {\bibfnamefont {N.}~\bibnamefont {Taberlet}}, \ and\
  \bibinfo {author} {\bibfnamefont {S.}~\bibnamefont {Manneville}},\ }\href
  {\doibase {10.1039/c3sm52548a}} {\bibfield  {journal} {\bibinfo  {journal}
  {{Soft Matter}}\ }\textbf {\bibinfo {volume} {{10}}},\ \bibinfo {pages}
  {1555} (\bibinfo {year} {{2014}})}\BibitemShut {NoStop}%
\bibitem [{\citenamefont {Gibaud}\ \emph {et~al.}(2009)\citenamefont {Gibaud},
  \citenamefont {Barentin}, \citenamefont {Taberlet},\ and\ \citenamefont
  {Manneville}}]{ISI:000268689400009}%
  \BibitemOpen
  \bibfield  {author} {\bibinfo {author} {\bibfnamefont {T.}~\bibnamefont
  {Gibaud}}, \bibinfo {author} {\bibfnamefont {C.}~\bibnamefont {Barentin}},
  \bibinfo {author} {\bibfnamefont {N.}~\bibnamefont {Taberlet}}, \ and\
  \bibinfo {author} {\bibfnamefont {S.}~\bibnamefont {Manneville}},\ }\href
  {\doibase {10.1039/b906274b}} {\bibfield  {journal} {\bibinfo  {journal}
  {{Soft Matter}}\ }\textbf {\bibinfo {volume} {{5}}},\ \bibinfo {pages} {3026}
  (\bibinfo {year} {{2009}})}\BibitemShut {NoStop}%
\bibitem [{\citenamefont {Gibaud}\ \emph {et~al.}(2010)\citenamefont {Gibaud},
  \citenamefont {Frelat},\ and\ \citenamefont
  {Manneville}}]{ISI:000280140800011}%
  \BibitemOpen
  \bibfield  {author} {\bibinfo {author} {\bibfnamefont {T.}~\bibnamefont
  {Gibaud}}, \bibinfo {author} {\bibfnamefont {D.}~\bibnamefont {Frelat}}, \
  and\ \bibinfo {author} {\bibfnamefont {S.}~\bibnamefont {Manneville}},\
  }\href {\doibase {10.1039/c000886a}} {\bibfield  {journal} {\bibinfo
  {journal} {{Soft Matter}}\ }\textbf {\bibinfo {volume} {{6}}},\ \bibinfo
  {pages} {3482} (\bibinfo {year} {{2010}})}\BibitemShut {NoStop}%
\bibitem [{\citenamefont {Gibaud}\ \emph {et~al.}(2008)\citenamefont {Gibaud},
  \citenamefont {Barentin},\ and\ \citenamefont
  {Manneville}}]{ISI:000261891200077}%
  \BibitemOpen
  \bibfield  {author} {\bibinfo {author} {\bibfnamefont {T.}~\bibnamefont
  {Gibaud}}, \bibinfo {author} {\bibfnamefont {C.}~\bibnamefont {Barentin}}, \
  and\ \bibinfo {author} {\bibfnamefont {S.}~\bibnamefont {Manneville}},\
  }\href {\doibase {10.1103/PhysRevLett.101.258302}} {\bibfield  {journal}
  {\bibinfo  {journal} {{Phys. Rev. Lett.}}\ }\textbf {\bibinfo {volume}
  {{101}}},\ \bibinfo {pages} {258302} (\bibinfo {year} {{2008}})}\BibitemShut
  {NoStop}%
\bibitem [{\citenamefont {Kurokawa}\ \emph {et~al.}(2015)\citenamefont
  {Kurokawa}, \citenamefont {Vidal}, \citenamefont {Kurita}, \citenamefont
  {Divoux},\ and\ \citenamefont {Manneville}}]{ISI:000365222200015}%
  \BibitemOpen
  \bibfield  {author} {\bibinfo {author} {\bibfnamefont {A.}~\bibnamefont
  {Kurokawa}}, \bibinfo {author} {\bibfnamefont {V.}~\bibnamefont {Vidal}},
  \bibinfo {author} {\bibfnamefont {K.}~\bibnamefont {Kurita}}, \bibinfo
  {author} {\bibfnamefont {T.}~\bibnamefont {Divoux}}, \ and\ \bibinfo {author}
  {\bibfnamefont {S.}~\bibnamefont {Manneville}},\ }\href {\doibase
  {10.1039/c5sm01259g}} {\bibfield  {journal} {\bibinfo  {journal} {{Soft
  Matter}}\ }\textbf {\bibinfo {volume} {{11}}},\ \bibinfo {pages} {9026}
  (\bibinfo {year} {{2015}})}\BibitemShut {NoStop}%
\bibitem [{\citenamefont {Sentjabrskaja}\ \emph {et~al.}(2015)\citenamefont
  {Sentjabrskaja}, \citenamefont {Chaudhuri}, \citenamefont {Hermes},
  \citenamefont {Poon}, \citenamefont {Horbach}, \citenamefont {Egelhaaf},\
  and\ \citenamefont {Laurati}}]{ISI:000357577400001}%
  \BibitemOpen
  \bibfield  {author} {\bibinfo {author} {\bibfnamefont {T.}~\bibnamefont
  {Sentjabrskaja}}, \bibinfo {author} {\bibfnamefont {P.}~\bibnamefont
  {Chaudhuri}}, \bibinfo {author} {\bibfnamefont {M.}~\bibnamefont {Hermes}},
  \bibinfo {author} {\bibfnamefont {W.~C.~K.}\ \bibnamefont {Poon}}, \bibinfo
  {author} {\bibfnamefont {J.}~\bibnamefont {Horbach}}, \bibinfo {author}
  {\bibfnamefont {S.~U.}\ \bibnamefont {Egelhaaf}}, \ and\ \bibinfo {author}
  {\bibfnamefont {M.}~\bibnamefont {Laurati}},\ }\href {\doibase
  {10.1038/srep11884}} {\bibfield  {journal} {\bibinfo  {journal} {{Sci.
  Rep.}}\ }\textbf {\bibinfo {volume} {{5}}},\ \bibinfo {pages} {11884}
  (\bibinfo {year} {{2015}})}\BibitemShut {NoStop}%
\bibitem [{\citenamefont {Schuh}\ \emph {et~al.}(2007)\citenamefont {Schuh},
  \citenamefont {Hufnagel},\ and\ \citenamefont
  {Ramamurty}}]{schuh2007mechanical}%
  \BibitemOpen
  \bibfield  {author} {\bibinfo {author} {\bibfnamefont {C.~A.}\ \bibnamefont
  {Schuh}}, \bibinfo {author} {\bibfnamefont {T.~C.}\ \bibnamefont {Hufnagel}},
  \ and\ \bibinfo {author} {\bibfnamefont {U.}~\bibnamefont {Ramamurty}},\
  }\href@noop {} {\bibfield  {journal} {\bibinfo  {journal} {Acta Materialia}\
  }\textbf {\bibinfo {volume} {55}},\ \bibinfo {pages} {4067} (\bibinfo {year}
  {2007})}\BibitemShut {NoStop}%
\bibitem [{\citenamefont {Jaiswal}\ \emph {et~al.}(2016)\citenamefont
  {Jaiswal}, \citenamefont {Procaccia}, \citenamefont {Rainone},\ and\
  \citenamefont {Singh}}]{ISI:000370815100008}%
  \BibitemOpen
  \bibfield  {author} {\bibinfo {author} {\bibfnamefont {P.~K.}\ \bibnamefont
  {Jaiswal}}, \bibinfo {author} {\bibfnamefont {I.}~\bibnamefont {Procaccia}},
  \bibinfo {author} {\bibfnamefont {C.}~\bibnamefont {Rainone}}, \ and\
  \bibinfo {author} {\bibfnamefont {M.}~\bibnamefont {Singh}},\ }\href
  {\doibase {10.1103/PhysRevLett.116.085501}} {\bibfield  {journal} {\bibinfo
  {journal} {{Phys. Rev. Lett.}}\ }\textbf {\bibinfo {volume} {{116}}},\
  \bibinfo {pages} {085501} (\bibinfo {year} {{2016}})}\BibitemShut {NoStop}%
\bibitem [{\citenamefont {Procaccia}\ \emph {et~al.}(2017)\citenamefont
  {Procaccia}, \citenamefont {Rainone},\ and\ \citenamefont
  {Singh}}]{ISI:000410885300004}%
  \BibitemOpen
  \bibfield  {author} {\bibinfo {author} {\bibfnamefont {I.}~\bibnamefont
  {Procaccia}}, \bibinfo {author} {\bibfnamefont {C.}~\bibnamefont {Rainone}},
  \ and\ \bibinfo {author} {\bibfnamefont {M.}~\bibnamefont {Singh}},\ }\href
  {\doibase {10.1103/PhysRevE.96.032907}} {\bibfield  {journal} {\bibinfo
  {journal} {{Phys. Rev. E}}\ }\textbf {\bibinfo {volume} {{96}}},\ \bibinfo
  {pages} {032907} (\bibinfo {year} {{2017}})}\BibitemShut {NoStop}%
\bibitem [{\citenamefont {Lin}\ \emph {et~al.}(2015)\citenamefont {Lin},
  \citenamefont {Gueudre}, \citenamefont {Rosso},\ and\ \citenamefont
  {Wyart}}]{ISI:000362909100023}%
  \BibitemOpen
  \bibfield  {author} {\bibinfo {author} {\bibfnamefont {J.}~\bibnamefont
  {Lin}}, \bibinfo {author} {\bibfnamefont {T.}~\bibnamefont {Gueudre}},
  \bibinfo {author} {\bibfnamefont {A.}~\bibnamefont {Rosso}}, \ and\ \bibinfo
  {author} {\bibfnamefont {M.}~\bibnamefont {Wyart}},\ }\href {\doibase
  {10.1103/PhysRevLett.115.168001}} {\bibfield  {journal} {\bibinfo  {journal}
  {{Phys. Rev. Lett.}}\ }\textbf {\bibinfo {volume} {{115}}},\ \bibinfo {pages}
  {168001} (\bibinfo {year} {{2015}})}\BibitemShut {NoStop}%
\bibitem [{\citenamefont {Liu}\ \emph {et~al.}(2018)\citenamefont {Liu},
  \citenamefont {Ferrero}, \citenamefont {Martens},\ and\ \citenamefont
  {Barrat}}]{liu2018creep}%
  \BibitemOpen
  \bibfield  {author} {\bibinfo {author} {\bibfnamefont {C.}~\bibnamefont
  {Liu}}, \bibinfo {author} {\bibfnamefont {E.~E.}\ \bibnamefont {Ferrero}},
  \bibinfo {author} {\bibfnamefont {K.}~\bibnamefont {Martens}}, \ and\
  \bibinfo {author} {\bibfnamefont {J.-L.}\ \bibnamefont {Barrat}},\
  }\href@noop {} {\bibfield  {journal} {\bibinfo  {journal} {Soft matter}\
  }\textbf {\bibinfo {volume} {14}},\ \bibinfo {pages} {8306} (\bibinfo {year}
  {2018})}\BibitemShut {NoStop}%
\bibitem [{\citenamefont {Shrivastav}\ \emph
  {et~al.}(2016{\natexlab{a}})\citenamefont {Shrivastav}, \citenamefont
  {Chaudhuri},\ and\ \citenamefont {Horbach}}]{ISI:000386386400004}%
  \BibitemOpen
  \bibfield  {author} {\bibinfo {author} {\bibfnamefont {G.~P.}\ \bibnamefont
  {Shrivastav}}, \bibinfo {author} {\bibfnamefont {P.}~\bibnamefont
  {Chaudhuri}}, \ and\ \bibinfo {author} {\bibfnamefont {J.}~\bibnamefont
  {Horbach}},\ }\href {\doibase {10.1103/PhysRevE.94.042605}} {\bibfield
  {journal} {\bibinfo  {journal} {{Phys. Rev. E}}\ }\textbf {\bibinfo {volume}
  {{94}}},\ \bibinfo {pages} {042605} (\bibinfo {year}
  {{2016}}{\natexlab{a}})}\BibitemShut {NoStop}%
\bibitem [{\citenamefont {Shrivastav}\ \emph
  {et~al.}(2016{\natexlab{b}})\citenamefont {Shrivastav}, \citenamefont
  {Chaudhuri},\ and\ \citenamefont {Horbach}}]{ISI:000384392300003}%
  \BibitemOpen
  \bibfield  {author} {\bibinfo {author} {\bibfnamefont {G.~P.}\ \bibnamefont
  {Shrivastav}}, \bibinfo {author} {\bibfnamefont {P.}~\bibnamefont
  {Chaudhuri}}, \ and\ \bibinfo {author} {\bibfnamefont {J.}~\bibnamefont
  {Horbach}},\ }\href {\doibase {10.1122/1.4959967}} {\bibfield  {journal}
  {\bibinfo  {journal} {{J. Rheol.}}\ }\textbf {\bibinfo {volume} {{60}}},\
  \bibinfo {pages} {835} (\bibinfo {year} {{2016}}{\natexlab{b}})}\BibitemShut
  {NoStop}%
\bibitem [{\citenamefont {Wisitsorasak}\ and\ \citenamefont
  {Wolynes}(2012)}]{ISI:000309611400031}%
  \BibitemOpen
  \bibfield  {author} {\bibinfo {author} {\bibfnamefont {A.}~\bibnamefont
  {Wisitsorasak}}\ and\ \bibinfo {author} {\bibfnamefont {P.~G.}\ \bibnamefont
  {Wolynes}},\ }\href {\doibase 10.1073/pnas.1214130109} {\bibfield  {journal}
  {\bibinfo  {journal} {Proceedings of the National Academy of Sciences}\
  }\textbf {\bibinfo {volume} {109}},\ \bibinfo {pages} {16068} (\bibinfo
  {year} {2012})}\BibitemShut {NoStop}%
\bibitem [{\citenamefont {Parisi}\ \emph {et~al.}(2017)\citenamefont {Parisi},
  \citenamefont {Procaccia}, \citenamefont {Rainone},\ and\ \citenamefont
  {Singh}}]{ISI:000402296700034}%
  \BibitemOpen
  \bibfield  {author} {\bibinfo {author} {\bibfnamefont {G.}~\bibnamefont
  {Parisi}}, \bibinfo {author} {\bibfnamefont {I.}~\bibnamefont {Procaccia}},
  \bibinfo {author} {\bibfnamefont {C.}~\bibnamefont {Rainone}}, \ and\
  \bibinfo {author} {\bibfnamefont {M.}~\bibnamefont {Singh}},\ }\href
  {\doibase {10.1073/pnas.1700075114}} {\bibfield  {journal} {\bibinfo
  {journal} {{Proc. Natl. Acad. Sci. USA}}\ }\textbf {\bibinfo {volume}
  {{114}}},\ \bibinfo {pages} {5577} (\bibinfo {year} {{2017}})}\BibitemShut
  {NoStop}%
\bibitem [{\citenamefont {Nandi}\ \emph {et~al.}(2014)\citenamefont {Nandi},
  \citenamefont {Biroli}, \citenamefont {Bouchaud}, \citenamefont {Miyazaki},\
  and\ \citenamefont {Reichman}}]{ISI:000346387700013}%
  \BibitemOpen
  \bibfield  {author} {\bibinfo {author} {\bibfnamefont {S.~K.}\ \bibnamefont
  {Nandi}}, \bibinfo {author} {\bibfnamefont {G.}~\bibnamefont {Biroli}},
  \bibinfo {author} {\bibfnamefont {J.-P.}\ \bibnamefont {Bouchaud}}, \bibinfo
  {author} {\bibfnamefont {K.}~\bibnamefont {Miyazaki}}, \ and\ \bibinfo
  {author} {\bibfnamefont {D.~R.}\ \bibnamefont {Reichman}},\ }\href {\doibase
  {10.1103/PhysRevLett.113.245701}} {\bibfield  {journal} {\bibinfo  {journal}
  {{Phys. Rev. Lett.}}\ }\textbf {\bibinfo {volume} {{113}}},\ \bibinfo {pages}
  {245701} (\bibinfo {year} {{2014}})}\BibitemShut {NoStop}%
\bibitem [{\citenamefont {Rainone}\ \emph {et~al.}(2015)\citenamefont
  {Rainone}, \citenamefont {Urbani}, \citenamefont {Yoshino},\ and\
  \citenamefont {Zamponi}}]{rainone2015following}%
  \BibitemOpen
  \bibfield  {author} {\bibinfo {author} {\bibfnamefont {C.}~\bibnamefont
  {Rainone}}, \bibinfo {author} {\bibfnamefont {P.}~\bibnamefont {Urbani}},
  \bibinfo {author} {\bibfnamefont {H.}~\bibnamefont {Yoshino}}, \ and\
  \bibinfo {author} {\bibfnamefont {F.}~\bibnamefont {Zamponi}},\ }\href@noop
  {} {\bibfield  {journal} {\bibinfo  {journal} {Physical review letters}\
  }\textbf {\bibinfo {volume} {114}},\ \bibinfo {pages} {015701} (\bibinfo
  {year} {2015})}\BibitemShut {NoStop}%
\bibitem [{\citenamefont {Urbani}\ and\ \citenamefont
  {Zamponi}(2017)}]{urbani2017shear}%
  \BibitemOpen
  \bibfield  {author} {\bibinfo {author} {\bibfnamefont {P.}~\bibnamefont
  {Urbani}}\ and\ \bibinfo {author} {\bibfnamefont {F.}~\bibnamefont
  {Zamponi}},\ }\href@noop {} {\bibfield  {journal} {\bibinfo  {journal}
  {Physical review letters}\ }\textbf {\bibinfo {volume} {118}},\ \bibinfo
  {pages} {038001} (\bibinfo {year} {2017})}\BibitemShut {NoStop}%
\bibitem [{\citenamefont {Aime}\ \emph {et~al.}(2018)\citenamefont {Aime},
  \citenamefont {Ramos},\ and\ \citenamefont
  {Cipelletti}}]{ISI:000429012500051}%
  \BibitemOpen
  \bibfield  {author} {\bibinfo {author} {\bibfnamefont {S.}~\bibnamefont
  {Aime}}, \bibinfo {author} {\bibfnamefont {L.}~\bibnamefont {Ramos}}, \ and\
  \bibinfo {author} {\bibfnamefont {L.}~\bibnamefont {Cipelletti}},\ }\href
  {\doibase {10.1073/pnas.1717403115}} {\bibfield  {journal} {\bibinfo
  {journal} {{Proc. Natl. Acad. Sci. USA}}\ }\textbf {\bibinfo {volume}
  {{115}}},\ \bibinfo {pages} {3587} (\bibinfo {year} {{2018}})}\BibitemShut
  {NoStop}%
\bibitem [{\citenamefont {Aime}\ \emph {et~al.}(2016)\citenamefont {Aime},
  \citenamefont {Ramos}, \citenamefont {Fromental}, \citenamefont {Prevot},
  \citenamefont {Jelinek},\ and\ \citenamefont
  {Cipelletti}}]{ISI:000392096800036}%
  \BibitemOpen
  \bibfield  {author} {\bibinfo {author} {\bibfnamefont {S.}~\bibnamefont
  {Aime}}, \bibinfo {author} {\bibfnamefont {L.}~\bibnamefont {Ramos}},
  \bibinfo {author} {\bibfnamefont {J.~M.}\ \bibnamefont {Fromental}}, \bibinfo
  {author} {\bibfnamefont {G.}~\bibnamefont {Prevot}}, \bibinfo {author}
  {\bibfnamefont {R.}~\bibnamefont {Jelinek}}, \ and\ \bibinfo {author}
  {\bibfnamefont {L.}~\bibnamefont {Cipelletti}},\ }\href {\doibase
  {10.1063/1.4972253}} {\bibfield  {journal} {\bibinfo  {journal} {{Rev. Sci.
  Instrum.}}\ }\textbf {\bibinfo {volume} {{87}}},\ \bibinfo {pages} {123907}
  (\bibinfo {year} {{2016}})}\BibitemShut {NoStop}%
\bibitem [{\citenamefont {Knowlton}\ \emph {et~al.}(2014)\citenamefont
  {Knowlton}, \citenamefont {Pine},\ and\ \citenamefont
  {Cipelletti}}]{ISI:000341025700007}%
  \BibitemOpen
  \bibfield  {author} {\bibinfo {author} {\bibfnamefont {E.~D.}\ \bibnamefont
  {Knowlton}}, \bibinfo {author} {\bibfnamefont {D.~J.}\ \bibnamefont {Pine}},
  \ and\ \bibinfo {author} {\bibfnamefont {L.}~\bibnamefont {Cipelletti}},\
  }\href {\doibase {10.1039/c4sm00531g}} {\bibfield  {journal} {\bibinfo
  {journal} {{Soft Matter}}\ }\textbf {\bibinfo {volume} {{10}}},\ \bibinfo
  {pages} {6931} (\bibinfo {year} {{2014}})}\BibitemShut {NoStop}%
\bibitem [{\citenamefont {Ozawa}\ \emph {et~al.}(2018)\citenamefont {Ozawa},
  \citenamefont {Berthier}, \citenamefont {Biroli}, \citenamefont {Rosso},\
  and\ \citenamefont {Tarjus}}]{ISI:000436245000061}%
  \BibitemOpen
  \bibfield  {author} {\bibinfo {author} {\bibfnamefont {M.}~\bibnamefont
  {Ozawa}}, \bibinfo {author} {\bibfnamefont {L.}~\bibnamefont {Berthier}},
  \bibinfo {author} {\bibfnamefont {G.}~\bibnamefont {Biroli}}, \bibinfo
  {author} {\bibfnamefont {A.}~\bibnamefont {Rosso}}, \ and\ \bibinfo {author}
  {\bibfnamefont {G.}~\bibnamefont {Tarjus}},\ }\href {\doibase
  {10.1073/pnas.1806156115}} {\bibfield  {journal} {\bibinfo  {journal}
  {{Proceedings of the National Academy of Sciences of the United States of
  America}}\ }\textbf {\bibinfo {volume} {{115}}},\ \bibinfo {pages} {6656}
  (\bibinfo {year} {{2018}})}\BibitemShut {NoStop}%
\bibitem [{\citenamefont {Popovi{\'c}}\ \emph {et~al.}(2018)\citenamefont
  {Popovi{\'c}}, \citenamefont {de~Geus},\ and\ \citenamefont
  {Wyart}}]{popovic2018elastoplastic}%
  \BibitemOpen
  \bibfield  {author} {\bibinfo {author} {\bibfnamefont {M.}~\bibnamefont
  {Popovi{\'c}}}, \bibinfo {author} {\bibfnamefont {T.~W.}\ \bibnamefont
  {de~Geus}}, \ and\ \bibinfo {author} {\bibfnamefont {M.}~\bibnamefont
  {Wyart}},\ }\href@noop {} {\bibfield  {journal} {\bibinfo  {journal}
  {Physical Review E}\ }\textbf {\bibinfo {volume} {98}},\ \bibinfo {pages}
  {040901} (\bibinfo {year} {2018})}\BibitemShut {NoStop}%
\bibitem [{\citenamefont {Moorcroft}\ and\ \citenamefont
  {Fielding}(2013)}]{ISI:000315141600016}%
  \BibitemOpen
  \bibfield  {author} {\bibinfo {author} {\bibfnamefont {R.~L.}\ \bibnamefont
  {Moorcroft}}\ and\ \bibinfo {author} {\bibfnamefont {S.~M.}\ \bibnamefont
  {Fielding}},\ }\href {\doibase {10.1103/PhysRevLett.110.086001}} {\bibfield
  {journal} {\bibinfo  {journal} {{Phys. Rev. Lett.}}\ }\textbf {\bibinfo
  {volume} {{110}}},\ \bibinfo {pages} {086001} (\bibinfo {year}
  {{2013}})}\BibitemShut {NoStop}%
\bibitem [{\citenamefont {Manning}\ \emph {et~al.}(2007)\citenamefont
  {Manning}, \citenamefont {Langer},\ and\ \citenamefont
  {Carlson}}]{ISI:000251326200013}%
  \BibitemOpen
  \bibfield  {author} {\bibinfo {author} {\bibfnamefont {M.~L.}\ \bibnamefont
  {Manning}}, \bibinfo {author} {\bibfnamefont {J.~S.}\ \bibnamefont {Langer}},
  \ and\ \bibinfo {author} {\bibfnamefont {J.~M.}\ \bibnamefont {Carlson}},\
  }\href {\doibase {10.1103/PhysRevE.76.056106}} {\bibfield  {journal}
  {\bibinfo  {journal} {{Phys. Rev. E}}\ }\textbf {\bibinfo {volume} {{76}}},\
  \bibinfo {pages} {056106} (\bibinfo {year} {{2007}})}\BibitemShut {NoStop}%
\bibitem [{\citenamefont {Manning}\ \emph {et~al.}(2009)\citenamefont
  {Manning}, \citenamefont {Daub}, \citenamefont {Langer},\ and\ \citenamefont
  {Carlson}}]{ISI:000262976900016}%
  \BibitemOpen
  \bibfield  {author} {\bibinfo {author} {\bibfnamefont {M.~L.}\ \bibnamefont
  {Manning}}, \bibinfo {author} {\bibfnamefont {E.~G.}\ \bibnamefont {Daub}},
  \bibinfo {author} {\bibfnamefont {J.~S.}\ \bibnamefont {Langer}}, \ and\
  \bibinfo {author} {\bibfnamefont {J.~M.}\ \bibnamefont {Carlson}},\ }\href
  {\doibase {10.1103/PhysRevE.79.016110}} {\bibfield  {journal} {\bibinfo
  {journal} {{Phys. Rev. E}}\ }\textbf {\bibinfo {volume} {{79}}},\ \bibinfo
  {pages} {016110} (\bibinfo {year} {{2009}})}\BibitemShut {NoStop}%
\bibitem [{\citenamefont {Moorcroft}\ \emph {et~al.}(2011)\citenamefont
  {Moorcroft}, \citenamefont {Cates},\ and\ \citenamefont
  {Fielding}}]{ISI:000286879900011}%
  \BibitemOpen
  \bibfield  {author} {\bibinfo {author} {\bibfnamefont {R.~L.}\ \bibnamefont
  {Moorcroft}}, \bibinfo {author} {\bibfnamefont {M.~E.}\ \bibnamefont
  {Cates}}, \ and\ \bibinfo {author} {\bibfnamefont {S.~M.}\ \bibnamefont
  {Fielding}},\ }\href {\doibase {10.1103/PhysRevLett.106.055502}} {\bibfield
  {journal} {\bibinfo  {journal} {{Phys. Rev. Lett.}}\ }\textbf {\bibinfo
  {volume} {{106}}},\ \bibinfo {pages} {055502} (\bibinfo {year}
  {{2011}})}\BibitemShut {NoStop}%
\bibitem [{\citenamefont {Nicolas}\ \emph {et~al.}(2018)\citenamefont
  {Nicolas}, \citenamefont {Ferrero}, \citenamefont {Martens},\ and\
  \citenamefont {Barrat}}]{nicolas2018deformation}%
  \BibitemOpen
  \bibfield  {author} {\bibinfo {author} {\bibfnamefont {A.}~\bibnamefont
  {Nicolas}}, \bibinfo {author} {\bibfnamefont {E.~E.}\ \bibnamefont
  {Ferrero}}, \bibinfo {author} {\bibfnamefont {K.}~\bibnamefont {Martens}}, \
  and\ \bibinfo {author} {\bibfnamefont {J.-L.}\ \bibnamefont {Barrat}},\
  }\href@noop {} {\bibfield  {journal} {\bibinfo  {journal} {Rev. Mod. Phys.}\
  }\textbf {\bibinfo {volume} {90}},\ \bibinfo {pages} {045006} (\bibinfo
  {year} {2018})}\BibitemShut {NoStop}%
\bibitem [{PRL()}]{PRL_SI}%
  \BibitemOpen
  \href@noop {} {}\bibinfo {note} {Supplementary material.}\BibitemShut {Stop}%
\bibitem [{\citenamefont {Sollich}\ \emph {et~al.}(1997)\citenamefont
  {Sollich}, \citenamefont {Lequeux}, \citenamefont {Hebraud},\ and\
  \citenamefont {Cates}}]{ISI:A1997WM06400048}%
  \BibitemOpen
  \bibfield  {author} {\bibinfo {author} {\bibfnamefont {P.}~\bibnamefont
  {Sollich}}, \bibinfo {author} {\bibfnamefont {F.}~\bibnamefont {Lequeux}},
  \bibinfo {author} {\bibfnamefont {P.}~\bibnamefont {Hebraud}}, \ and\
  \bibinfo {author} {\bibfnamefont {M.}~\bibnamefont {Cates}},\ }\href
  {\doibase {10.1103/PhysRevLett.78.2020}} {\bibfield  {journal} {\bibinfo
  {journal} {{Phys. Rev. Lett.}}\ }\textbf {\bibinfo {volume} {{78}}},\
  \bibinfo {pages} {2020} (\bibinfo {year} {{1997}})}\BibitemShut {NoStop}%
\bibitem [{\citenamefont {Vasoya}\ \emph {et~al.}(2016)\citenamefont {Vasoya},
  \citenamefont {Rycroft},\ and\ \citenamefont
  {Bouchbinder}}]{vasoya2016notch}%
  \BibitemOpen
  \bibfield  {author} {\bibinfo {author} {\bibfnamefont {M.}~\bibnamefont
  {Vasoya}}, \bibinfo {author} {\bibfnamefont {C.~H.}\ \bibnamefont {Rycroft}},
  \ and\ \bibinfo {author} {\bibfnamefont {E.}~\bibnamefont {Bouchbinder}},\
  }\href@noop {} {\bibfield  {journal} {\bibinfo  {journal} {Physical Review
  Applied}\ }\textbf {\bibinfo {volume} {6}},\ \bibinfo {pages} {024008}
  (\bibinfo {year} {2016})}\BibitemShut {NoStop}%
\bibitem [{\citenamefont {Ketkaew}\ \emph {et~al.}(2018)\citenamefont
  {Ketkaew}, \citenamefont {Chen}, \citenamefont {Wang}, \citenamefont {Datye},
  \citenamefont {Fan}, \citenamefont {Pereira}, \citenamefont {Schwarz},
  \citenamefont {Liu}, \citenamefont {Yamada}, \citenamefont {Dmowski} \emph
  {et~al.}}]{ketkaew2018mechanical}%
  \BibitemOpen
  \bibfield  {author} {\bibinfo {author} {\bibfnamefont {J.}~\bibnamefont
  {Ketkaew}}, \bibinfo {author} {\bibfnamefont {W.}~\bibnamefont {Chen}},
  \bibinfo {author} {\bibfnamefont {H.}~\bibnamefont {Wang}}, \bibinfo {author}
  {\bibfnamefont {A.}~\bibnamefont {Datye}}, \bibinfo {author} {\bibfnamefont
  {M.}~\bibnamefont {Fan}}, \bibinfo {author} {\bibfnamefont {G.}~\bibnamefont
  {Pereira}}, \bibinfo {author} {\bibfnamefont {U.~D.}\ \bibnamefont
  {Schwarz}}, \bibinfo {author} {\bibfnamefont {Z.}~\bibnamefont {Liu}},
  \bibinfo {author} {\bibfnamefont {R.}~\bibnamefont {Yamada}}, \bibinfo
  {author} {\bibfnamefont {W.}~\bibnamefont {Dmowski}},  \emph {et~al.},\
  }\href@noop {} {\bibfield  {journal} {\bibinfo  {journal} {Nature
  communications}\ }\textbf {\bibinfo {volume} {9}},\ \bibinfo {pages} {1}
  (\bibinfo {year} {2018})}\BibitemShut {NoStop}%
\bibitem [{\citenamefont {Singh}\ \emph {et~al.}(2020)\citenamefont {Singh},
  \citenamefont {Ozawa},\ and\ \citenamefont {Berthier}}]{singh2020brittle}%
  \BibitemOpen
  \bibfield  {author} {\bibinfo {author} {\bibfnamefont {M.}~\bibnamefont
  {Singh}}, \bibinfo {author} {\bibfnamefont {M.}~\bibnamefont {Ozawa}}, \ and\
  \bibinfo {author} {\bibfnamefont {L.}~\bibnamefont {Berthier}},\ }\href@noop
  {} {\bibfield  {journal} {\bibinfo  {journal} {Physical Review Materials}\
  }\textbf {\bibinfo {volume} {4}},\ \bibinfo {pages} {025603} (\bibinfo {year}
  {2020})}\BibitemShut {NoStop}%
\end{thebibliography}
\end{document}